%% file: main.tex
\documentclass[sigconf]{acmart}

\setcopyright{none}

\renewcommand{\footnotetextcopyrightpermission}[1]{}  
\renewcommand{\acmSubmissionID}{}  

\usepackage{float}
\usepackage{graphicx}
\usepackage{subcaption}
\usepackage{multirow}
\usepackage{pifont}
\usepackage{enumitem}
\usepackage{xspace}
\usepackage{colortbl}

\newcommand\vldbdoi{XX.XX/XXX.XX}
\newcommand\vldbpages{XXX-XXX}
\newcommand\vldbvolume{14}
\newcommand\vldbissue{1}
\newcommand\vldbyear{2020}
\newcommand\vldbauthors{\authors}
\newcommand\vldbtitle{\shorttitle} 
\newcommand\vldbavailabilityurl{URL_TO_YOUR_ARTIFACTS}
\newcommand\vldbpagestyle{plain}

\newcommand{\zw}[1]{\textcolor{purple}{#1}}
\newcommand{\zxh}[1]{\textcolor{blue}{#1}}

\newcommand{\blue}[1]{\textcolor{blue}{#1}}
\newcommand{\gray}[1]{\textcolor{gray}{#1}}

\newcommand{\hi}[1]{\vspace{.25em} \noindent {\bf #1} }
\newcommand{\llm}{\textsc{LLM}\xspace}
\newcommand{\llms}{\textsc{LLMs}\xspace}
\newcommand{\bfit}[1]{\textbf{\textit{#1}}}
\newcommand{\oursys}{\texttt{DBAIOps}\xspace}

\newcommand{\anomalymodel}{\texttt{AnomalyModel}\xspace}

\newcommand{\experiencegraph}{\texttt{ExperienceGraph}\xspace}
\newcommand{\anomalyprocessor}{\texttt{AnomalyProcessor}\xspace}
\newcommand{\experienceretriever}{\texttt{ExperienceRetriever}\xspace}
\newcommand{\rootcauseanalyser}{\texttt{RootCauseAnlyser}\xspace}

\newcommand{\om}{{O\&M}\xspace}

\newtheorem{definition}{DEFINITION}[section]

\begin{document}

\title{\oursys: Harnessing Decades of Expert Knowledge for Scalable Database Maintenance Powered by LLMs}
\title{\oursys: Deploying Knowledge-Graph-Driven Database Maintenance System with Long-Term Reasoning LLMs}
\title{DBAIOps: A Reasoning LLM-Enhanced Database Operation and Maintenance System using Knowledge Graphs}






\settopmatter{authorsperrow=4}

\pagestyle{plain}
\pagenumbering{arabic}

\author{Wei Zhou}
\affiliation{
    Shanghai Jiao Tong University
}
\email{weizhoudb@sjtu.edu.cn}

\author{Peng Sun}
\affiliation{
    Baisheng (Shenzhen) Technology Co., Ltd.
}
\email{sunpeng@dbaiops.com}

\author{Xuanhe Zhou}
\affiliation{
    Shanghai Jiao Tong University
}
\email{zhouxh@cs.sjtu.edu.cn}

\author{Qianglei Zang}
\affiliation{
    Baisheng (Shenzhen) Technology Co., Ltd.
}
\email{zangqianglei@dbaiops.com}

\author{Ji Xu}
\affiliation{
    Baisheng (Shenzhen) Technology Co., Ltd.
}
\email{xuji@dbaiops.com}

\author{Tieying Zhang}
\affiliation{
    \institution{Bytedance}
    \city{tieying.zhang}
}
\email{@bytedance.com}

\author{Guoliang Li}
\affiliation{
    \institution{Tsinghua University}
    \city{liguoliang}
}
\email{@tsinghua.edu.cn}

\author{Fan Wu}
\affiliation{
    Shanghai Jiao Tong University
}
\email{fwu@cs.sjtu.edu.cn}

\renewcommand{\shortauthors}{Zhou et al.}

\begin{abstract}
The operation and maintenance (O\&M) of database systems is critical to ensuring system availability and performance, typically requiring expert experience (e.g., identifying metric-to-anomaly relations) for effective diagnosis and recovery.  However, existing automatic database \om methods, including commercial products, cannot effectively utilize expert experience. On the one hand, rule-based methods only support basic \om tasks (e.g., metric-based anomaly detection), which are mostly numerical equations and cannot effectively incorporate literal \om experience (e.g., troubleshooting guidance in manuals). 
On the other hand, \llm-based methods, which retrieve fragmented information (e.g., standard documents + RAG), often generate inaccurate or generic results.

To address these limitations, we present \oursys, a novel hybrid database O\&M system that {\it combines reasoning LLMs with knowledge graphs} to achieve DBA-style diagnosis. 
First, \oursys introduces a heterogeneous graph model for representing the diagnosis experience, and proposes a semi-automatic graph construction algorithm to build that graph from {thousands of} documents. 
Second, \oursys develops a collection of (800+) reusable anomaly models that identify both directly alerted metrics and {implicitly correlated experience and metrics}. Third, for any given anomaly, \oursys employs an automatic graph evolution mechanism that explores the relevant paths over the graph and dynamically explores potential gaps (missing paths) without human intervention. Based on the explored diagnosis paths, \oursys leverages reasoning \llm (e.g., DeepSeek-R1) that inputs the relevant pathways, identifies root causes, and generates clear {diagnosis reports} for both DBAs and common users. Our evaluation over {four} mainstream database systems (Oracle, MySQL, PostgreSQL, and {DM8}) demonstrates that \oursys outperforms state-of-the-art baselines, {34.85\% and 47.22\% higher in root cause and human evaluation accuracy, respectively}.
\oursys supports {25} database systems and has been deployed in 20 real-world scenarios, covering domains like finance, energy, and healthcare ({\it \textcolor{blue}{\url{https://www.dbaiops.com}}}).
\end{abstract}

\maketitle

\setcounter{table}{0}

\input{1_introduction}

\input{2_definition}
\input{3_overview}

\input{5_knowledge}

\input{4_observability}

\input{diagnosis}

\input{6_reasoning}

\input{7_experiment}
\input{9_conclusion}

\clearpage
\bibliographystyle{ACM-Reference-Format}
\balance
\bibliography{sample-cleaned}



\end{document}

%% file: 1_introduction.tex


\section{Introduction}
\label{sec:intro}



Database operation and maintenance (\om) aims to detect, analyze, and resolve various anomalies that arise in target database instances, which is of great importance to meet the rigorous requirements during the online usage of these instances, such as high availability (e.g., {achieving 99.99\% four nines availability with less than 52.6 minutes of downtime per year for critical services such as financial and e‑commerce systems}~\cite{downtime}) and performance (e.g., service-level agreements (SLAs) enforced by cloud service providers~\cite{Azure, AmazonDatabase, TRAP}). {For instance, the NOTAM database outage (an honest mistake that cost the country millions) resulted in over 10,000 flight delays and more than 1,300 cancellations}~\cite{FAA1,FAA2}.



\begin{figure}[!t]
  \centering
  \includegraphics[width=\linewidth]{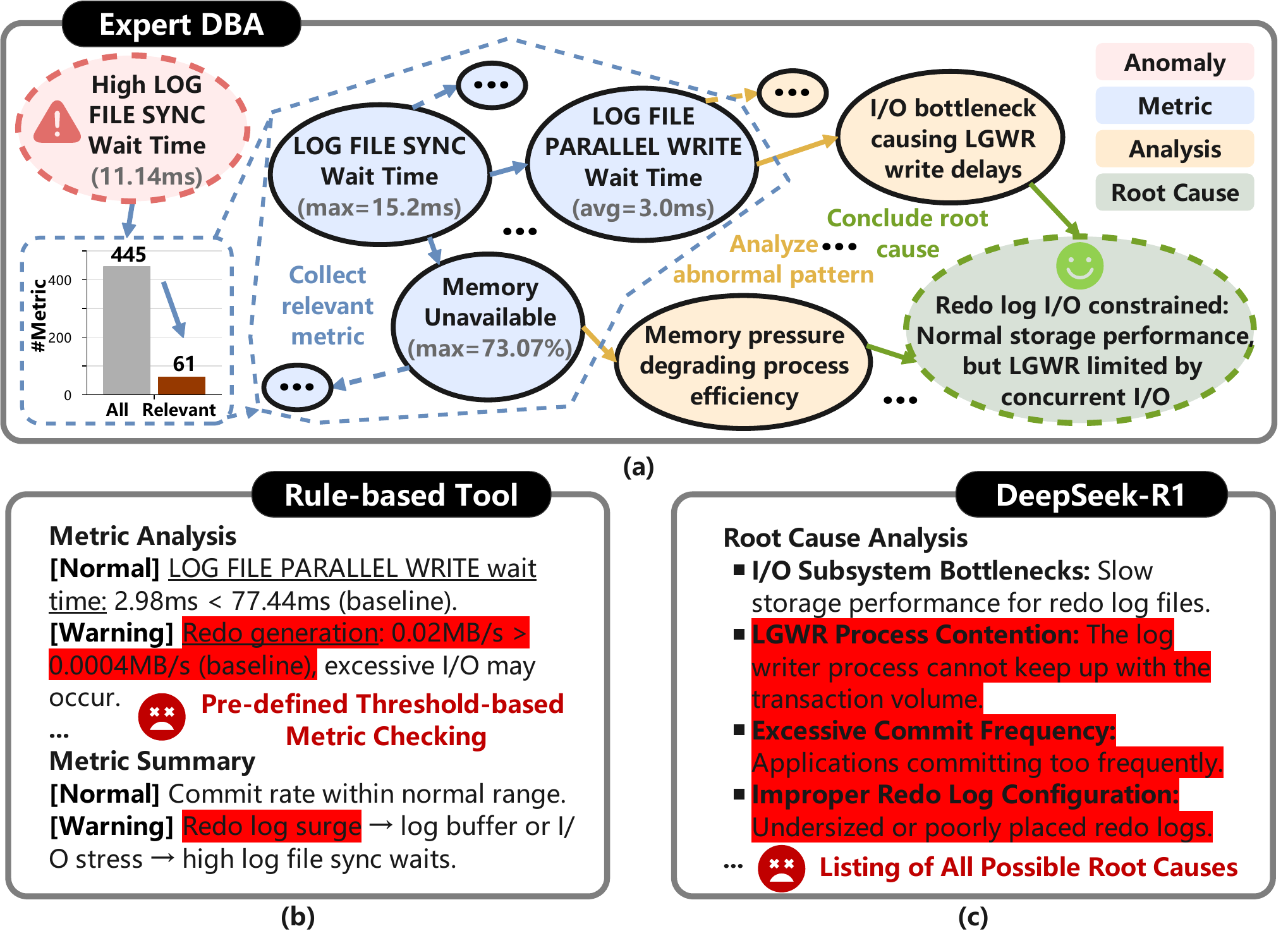}
  \vspace{-.65cm}
  \caption{Automatic database O\&M is challenging \textnormal{- (a) Expert DBA needs to analyze diverse information from triggered anomalies. (b) Empirical O\&M may apply misleading rules (caused by incorrect thresholds). (c) LLMs may lack O\&M experience and fail to diagnose {\it even with necessary abnormal information like relevant metrics.}}}
  \label{fig:intro}
  \vspace{-.7cm}
\end{figure}

Therefore, to ensure high availability and performance, many companies hire senior DBAs (with decades of experience) or purchase costly manual maintenance services~\cite{Rackspace,Percona,dataVail}.
For instance, as shown in Figure~\ref{fig:intro}, diagnosing anomalies such as \texttt{LOG\_FILE\_SYNC} traditionally relies on experienced DBAs to implicitly associate the alert with relevant metrics (e.g., \emph{LOG FILE PARALLEL WRITE Wait Time}) and construct causality chains to identify the root cause (e.g., ``I/O bottleneck limiting the LGWR process'').
It also involves filtering out large volumes of irrelevant metrics (e.g., selecting 61 relevant metrics from a total of 445), which could otherwise lead to incorrect conclusions.
However, this human-intensive \om process is time-consuming, difficult to scale, and becomes particularly inefficient when repeatedly applied to recurring anomalies.

\begin{sloppypar}
Existing methods leverage empirical rules~\cite{ADDM,ADPP} or even large language models (\llms)~\cite{panda,dbot2024, OurLLMSurvey} to automate some tasks in database \om (see Table~\ref{tab:relatedwork}). However, they do not support flexible DBA experience integration and demonstrate significant limitations. For instance, in  Figure~\ref{fig:intro}, rule-based methods like~\cite{ADDM,ADPP} integrate limited anomaly diagnosis rules with fixed thresholds (e.g., attributing the anomaly to a \texttt{Red\_log\_surge} merely because a preset threshold was exceeded). By contrast, \llm-based methods can infer subtler root causes by reasoning over system metrics and leveraging external knowledge~\cite{dbot2024}. However, they have two main problems, causing them hardly applicable in real scenarios. First, they have relatively low accuracy in matching relevant expert experience, which (1) naively chunk documents without capturing implicit relationships~\cite{panda}, and (2) rely solely on vector-based similarity for semantic matching~\cite{Andromeda}. Second, the retrieved experience is often fragmented, making it difficult for \llms to perform accurate step-by-step root cause analysis.
For instance, as shown in Figure~\ref{fig:intro}, DeepSeek-R1 lists multiple possible causes but fails to deliver a concrete, actionable diagnosis, such as identifying the exact root causes (e.g., Redo log I/O bottlenecks) or suggesting concrete recovery actions (e.g., optimizing Redo log placement).




\end{sloppypar}

To bridge the gap between experience-based \om practices of expert DBAs and the limited capabilities of existing methods, there are three main challenges.



\noindent \bfit{C1: How to effectively characterize and integrate \om experience?}
While extensive database \om experience exists across technical notes, scripts, and incident reports (mainly from large database companies~\cite{AmazonDatabase,AlibabaDatabase}), it is often fragmented and distributed in heterogeneous formats (e.g., informal textual documentation, unstructured log entries, and isolated SQL scripts). {\it There lacks an effective way to represent \om experience}, hindering their use in guiding \om tasks and supporting accurate database diagnosis (e.g., {correlating recurring \texttt{LOG\_FILE\_SYNC} wait events with storage I/O latency anomalies identified in historical incidents}).


\noindent \bfit{C2: How to capture implicitly correlated factors for diverse anomalies?} Effective analysis of database anomalies often necessitates a global view of system, log, and trace metrics (e.g., {redo log generation rate, I/O subsystem latency, and transaction commit frequency}). However, most existing methods focus primarily on metrics that exhibit abnormal patterns (e.g., changes in temporal trends or short-term fluctuations~\cite{CauseRank,Time1,Time2}). As a result, metrics that are {\it implicitly relevant but do not exhibit noticeable abnormal patterns} are frequently overlooked, leading to significant diagnosis errors. For instance, a normal log buffer hit ratio during \texttt{LOG\_FILE\_SYNC} waits may cause DBAs to overlook I/O bottleneck, only focusing on spiking sync wait times.

\noindent \bfit{C3: How to adaptively explore potential \om diagnosis paths?} 
Existing diagnosis systems often rely on fixed rule-based methods, which follow predefined decision paths and show limited adaptability over diverse scenarios~\cite{ADPP,ADDM}.
In addition, many of these systems either cannot generate informative reports that include root causes and promising recovery solutions~\cite{DBSherlock,ATDS,DBMind,iSQUAD}, or they generate overly technical outputs, such as exhaustive lists of metric values, that are difficult for common users to interpret and act upon~\cite{RCRank,CauseRank,PinSQL}. 

\begin{sloppypar} 
To address these challenges, we design \oursys, an experience-centric database \om~system integrating four key components:
(1) \emph{Heterogeneous \om Graph Model} for integrating the complicated and mostly-textual \om experience. This graph represents diagnosis paths through interconnected vertices and edges, {enabling the structured organization and incremental enrichment of \om experience for more precise diagnosis} \textbf{(for C1)};
(2) \emph{Correlation-Aware Anomaly Models} for identifying correlated factors associated with input anomaly. Each model incorporates statistical multi-metric correlation analysis, frequency control, and low-code tools to support the discovery of implicitly correlated metrics \textbf{(for C2)}; 
(3) \emph{Two-Stage Experience Retrieval Strategy} that adaptively traverses the graph to collect relevant diagnosis information (e.g., abnormal metrics), such that determining the correct diagnosis paths for different anomalies; 
(4) \emph{In-Context Reasoning-\llm Learning} for prompting the model to reason over the collected graph information and generate clear diagnosis reports that include both detailed root cause analysis and practical recovery solutions \textbf{(for C3)}.
\end{sloppypar}

\hi{Contributions.} We make the following contributions.

\noindent$\bullet$ We design a database operation and maintenance (O\&M) system for diagnosing real-world anomalies. {\it To the best of our knowledge, this is the first database O\&M system that integrates knowledge graph with reasoning \llms to identify root causes and provide recovery solutions} (see Section~\ref{sec:overview}).




\noindent$\bullet$ We propose a {graph-based experience model} to represent \om experience in graph paths. {\it Current experience model (the knowledge graph) comprises over 2,000 vertices and 800+ anomaly scenarios for {25} different database systems} (see Section~\ref{sec:graph} \& {\it \textcolor{blue}{\url{https://www.dbaiops.com}}}).


\noindent$\bullet$ We propose a {correlation-aware anomaly model} to capture implicit correlations across metrics and real-world anomalies, which can trigger {more accurate} graph exploration during online diagnosis (see Section~\ref{sec:anomaly}).

\noindent$\bullet$ We introduce a {two-stage graph evolution mechanism} that adaptively explores possible diagnosis paths for different anomalies. And we prompt \llm to reason over these diagnosis paths and generate diagnosis reports with specific recovery solutions (see Section~\ref{sec:diagnosis}). 

\noindent$\bullet$ Extensive experiments and case studies show that \oursys~outperforms both rule and \llm-based baselines in root cause accuracy ({34.85\% higher}) and human evaluation accuracy (47.22\% higher).

%% file: 2_definition.tex
\section{Background and Related Work}
\label{subsec:relatedwork}


Database \om refers to the process of maintaining and optimizing database systems, which typically involves (1) the collection of necessary \om factors (e.g., system metrics, logs, and traces) and (2) root cause diagnosis and recovery.  As shown in Table~\ref{tab:relatedwork}, we classify existing database \om methods into three main categories:



\noindent $\blacktriangleright$ {\bfit{Rule-based Methods.}} Methods in this category rely on human experts to incorporate their maintenance knowledge as rules into the diagnosis process, such as by defining a set of diagnosis paths for different types of anomalies~\cite{ADDM,DBSherlock,ATDS}. 
ADDM~\cite{ADDM} performs root cause diagnosis in a time graph based on rules (e.g. ``exploring all child nodes when a node's time is abnormal'').
DBSherlock~\cite{DBSherlock} encodes domain knowledge into rules and uses these rules to filter out predicates ($\mathit{Attr}>\mathit{k}$) that reflect secondary symptoms.
ADTS~\cite{ATDS} builds an expert system containing 175 rules in the form of ``Expression-Result'' statements to diagnose root causes. 

However, rule-based methods require specialized expertise to design and implement, and are generally limited to specific database systems. For instance, ADDM is only applicable to Oracle database, and extending it to other systems requires considerable manual efforts (e.g., incorporating new rules). In addition, the reliance on pre-defined rules and the absence of external knowledge integration reduces flexibility, making it difficult to adapt to new anomalies. 

\noindent $\blacktriangleright$ {\bfit{ML-based Methods.}} 
Methods in this category incorporate machine learning algorithms or models to enhance the root cause analysis accuracy of rule-based methods.
{CauseRank~\cite{CauseRank} employs a Bayesian Network Structure algorithm and expert rules to construct a causal graph of anomalies.}
{DBMind~\cite{DBMind} employs an LSTM-based encoder model to encode data into anomaly vectors for matching the root cause.}
iSQUAD~\cite{iSQUAD} employs the Bayesian Case Model to extract the key features of SQLs, while PinSQL\cite{PinSQL} employs an ML-based clustering algorithm to group SQLs according to their historical execution trends for root cause SQL diagnosis.
RCRank~\cite{RCRank} trains a multi-modal machine learning model to extract features from four types of data (SQL, log, plan, and metric) to rank the root causes of slow queries. 

However, since ML-based methods are typically built on top of rule-based systems, they inherit similar limitations. 
Moreover, ML models typically have poor generalization ability due to their strong dependence on training data~\cite{DeepLearning}, making them effective only for certain anomaly diagnosis.
For example, iSQUAD~\cite{iSQUAD} and PinSQL~\cite{PinSQL} are designed to diagnose slow SQLs of limited types.






\begin{table}[!t]
\caption{Comparison of Database \om Methods.}
\label{tab:relatedwork}
\resizebox{\linewidth}{!}{
\begin{tabular}{clccccc}
\rowcolor[HTML]{000000} 
\cellcolor[HTML]{000000}{\color[HTML]{FFFFFF} }                                    & \multicolumn{1}{c}{\cellcolor[HTML]{000000}{\color[HTML]{FFFFFF} }}                                  & \multicolumn{2}{c}{\cellcolor[HTML]{000000}{\color[HTML]{FFFFFF} \textbf{Diagnosis Evidence}}} & \cellcolor[HTML]{000000}{\color[HTML]{FFFFFF} }                                                                                            & \cellcolor[HTML]{000000}{\color[HTML]{FFFFFF} }                                                                                          & \cellcolor[HTML]{000000}{\color[HTML]{FFFFFF} }                                                                                         \\
\rowcolor[HTML]{000000} 
\multirow{-2}{*}{\cellcolor[HTML]{000000}{\color[HTML]{FFFFFF} \textbf{Category}}} & \multicolumn{1}{c}{\multirow{-2}{*}{\cellcolor[HTML]{000000}{\color[HTML]{FFFFFF} \textbf{Method}}}} & {\color[HTML]{FFFFFF} \textbf{Numeric}}         & {\color[HTML]{FFFFFF} \textbf{Text}}         & \multirow{-2}{*}{\cellcolor[HTML]{000000}{\color[HTML]{FFFFFF} \textbf{\begin{tabular}[c]{@{}c@{}}Experience\\ Integration\end{tabular}}}} & \multirow{-2}{*}{\cellcolor[HTML]{000000}{\color[HTML]{FFFFFF} \textbf{\begin{tabular}[c]{@{}c@{}}Experience\\ Evolution\end{tabular}}}} & \multirow{-2}{*}{\cellcolor[HTML]{000000}{\color[HTML]{FFFFFF} \textbf{\begin{tabular}[c]{@{}c@{}}New Anomaly\\ Support\end{tabular}}}} \\ \hline
\multicolumn{1}{|c|}{}                                                             & \multicolumn{1}{l|}{\textbf{ADDM~\cite{ADDM}}}                                                                   & \multicolumn{1}{c|}{\checkmark}                          & \multicolumn{1}{c|}{$\times$}                       & \multicolumn{1}{c|}{$\times$}                                                                                                                     & \multicolumn{1}{c|}{$\times$}                                                                                                                   & \multicolumn{1}{c|}{$\times$}                                                                                                                  \\ \cline{2-7} 
\multicolumn{1}{|c|}{}                                                             & \multicolumn{1}{l|}{\textbf{DBSherlock~\cite{DBSherlock}}}                                                             & \multicolumn{1}{c|}{\checkmark}                          & \multicolumn{1}{c|}{$\times$}                       & \multicolumn{1}{c|}{$\times$}                                                                                                                     & \multicolumn{1}{c|}{$\times$}                                                                                                                   & \multicolumn{1}{c|}{$\times$}                                                                                                                  \\ \cline{2-7} 
\multicolumn{1}{|c|}{\multirow{-3}{*}{\textbf{Rule-based}}}                        & \multicolumn{1}{l|}{\textbf{ADTS~\cite{ADPP}}}                                                                   & \multicolumn{1}{c|}{\checkmark}                          & \multicolumn{1}{c|}{$\times$}                       & \multicolumn{1}{c|}{$\times$}                                                                                                                     & \multicolumn{1}{c|}{$\times$}                                                                                                                   & \multicolumn{1}{c|}{$\times$}                                                                                                                  \\ \hline
\multicolumn{1}{|c|}{}                                                             & \multicolumn{1}{l|}{\textbf{CauseRank~\cite{CauseRank}}}                                                              & \multicolumn{1}{c|}{\checkmark}                          & \multicolumn{1}{c|}{$\times$}                       & \multicolumn{1}{c|}{$\times$}                                                                                                                     & \multicolumn{1}{c|}{$\times$}                                                                                                                   & \multicolumn{1}{c|}{$\times$}                                                                                                                  \\ \cline{2-7} 
\multicolumn{1}{|c|}{}                                                             & \multicolumn{1}{l|}{\textbf{DBMind~\cite{DBMind}}}                                                                 & \multicolumn{1}{c|}{\checkmark}                          & \multicolumn{1}{c|}{$\times$}                       & \multicolumn{1}{c|}{$\times$}                                                                                                                     & \multicolumn{1}{c|}{$\times$}                                                                                                                   & \multicolumn{1}{c|}{$\times$}                                                                                                                  \\ \cline{2-7} 
\multicolumn{1}{|c|}{}                                                             & \multicolumn{1}{l|}{\textbf{iSQUAD~\cite{iSQUAD}}}                                                                 & \multicolumn{1}{c|}{\checkmark}                          & \multicolumn{1}{c|}{$\times$}                       & \multicolumn{1}{c|}{$\times$}                                                                                                                     & \multicolumn{1}{c|}{$\times$}                                                                                                                   & \multicolumn{1}{c|}{$\times$}                                                                                                                  \\ \cline{2-7} 
\multicolumn{1}{|c|}{}                                                             & \multicolumn{1}{l|}{\textbf{PinSQL~\cite{PinSQL}}}                                                                 & \multicolumn{1}{c|}{\checkmark}                          & \multicolumn{1}{c|}{$\times$}                       & \multicolumn{1}{c|}{$\times$}                                                                                                                     & \multicolumn{1}{c|}{$\times$}                                                                                                                   & \multicolumn{1}{c|}{$\times$}                                                                                                                  \\ \cline{2-7} 
\multicolumn{1}{|c|}{\multirow{-5}{*}{\textbf{ML-based}}}                          & \multicolumn{1}{l|}{\textbf{RCRank~\cite{RCRank}}}                                                                 & \multicolumn{1}{c|}{\checkmark}                          & \multicolumn{1}{c|}{$\times$}                       & \multicolumn{1}{c|}{$\times$}                                                                                                                     & \multicolumn{1}{c|}{$\times$}                                                                                                                   & \multicolumn{1}{c|}{$\times$}                                                                                                                  \\ \hline
\multicolumn{1}{|c|}{}                                                             & \multicolumn{1}{l|}{\textbf{D-Bot~\cite{dbot2024}}}                                                                  & \multicolumn{1}{c|}{\checkmark}                          & \multicolumn{1}{c|}{\checkmark}                       & \multicolumn{1}{c|}{\checkmark}                                                                                                                     & \multicolumn{1}{c|}{$\times$}                                                                                                                   & \multicolumn{1}{c|}{\checkmark}                                                                                                                  \\ \cline{2-7} 
\multicolumn{1}{|c|}{}                                                             & \multicolumn{1}{l|}{\textbf{Panda~\cite{panda}}}                                                                  & \multicolumn{1}{c|}{$\times$}                          & \multicolumn{1}{c|}{\checkmark}                       & \multicolumn{1}{c|}{\checkmark}                                                                                                                     & \multicolumn{1}{c|}{$\times$}                                                                                                                   & \multicolumn{1}{c|}{\checkmark}                                                                                                                  \\ \cline{2-7} 
\multicolumn{1}{|c|}{}                                                             & \multicolumn{1}{l|}{\textbf{ChatDBA~\cite{chatdba}}}                                                                & \multicolumn{1}{c|}{$\times$}                          & \multicolumn{1}{c|}{\checkmark}                       & \multicolumn{1}{c|}{\checkmark}                                                                                                                     & \multicolumn{1}{c|}{$\times$}                                                                                                                   & \multicolumn{1}{c|}{\checkmark}                                                                                                                  \\ \cline{2-7} 
\multicolumn{1}{|c|}{}                                                             & \multicolumn{1}{l|}{\textbf{Andromeda~\cite{Andromeda}}}                                                              & \multicolumn{1}{c|}{$\times$}                          & \multicolumn{1}{c|}{\checkmark}                       & \multicolumn{1}{c|}{\checkmark}                                                                                                                     & \multicolumn{1}{c|}{$\times$}                                                                                                                   & \multicolumn{1}{c|}{\checkmark}                                                                                                                  \\ \cline{2-7} 
\multicolumn{1}{|c|}{}                                                             & \multicolumn{1}{l|}{\textbf{GaussMaster~\cite{gaussmaster}}}                                                            & \multicolumn{1}{c|}{\checkmark}                          & \multicolumn{1}{c|}{\checkmark}                       & \multicolumn{1}{c|}{\checkmark}                                                                                                                     & \multicolumn{1}{c|}{$\times$}                                                                                                                   & \multicolumn{1}{c|}{\checkmark}                                                                                                                  \\ \cline{2-7} 
\multicolumn{1}{|c|}{\multirow{-6}{*}{\textbf{LLM-based}}}                         & \multicolumn{1}{l|}{\textbf{\oursys}}                                                                & \multicolumn{1}{c|}{\checkmark}                          & \multicolumn{1}{c|}{\checkmark}                       & \multicolumn{1}{c|}{\checkmark}                                                                                                                     & \multicolumn{1}{c|}{\checkmark}                                                                                                                   & \multicolumn{1}{c|}{\checkmark}                                                                                                                  \\ \hline
\end{tabular}
}
\vspace{-2em}
\end{table}

\noindent {$\blacktriangleright$ \bfit{LLM-based Methods.}}
Methods in this category leverage the comprehension and reasoning capabilities of \llms to improve diagnosis accuracy and adaptability.
These methods utilize both the \llm’s internal knowledge (e.g., general understanding of different database systems) and external resources (e.g., historical anomaly cases).
For example, D-Bot~\cite{dbot2024} empowers \llm to perform diagnosis with prompts generated with matched document knowledge and retrieved tools and conduct multi-step root cause analysis using the tree-search-based algorithm.
ChatDBA~\cite{chatdba} leverages a decision tree structure to retrieve relevant information and instruct \llm-driven diagnosis.
{Panda~\cite{panda} and GaussMaster~\cite{gaussmaster} utilize \llm agents to specialized diagnosis modules or expert roles for collaborative diagnosis.
Andromeda~\cite{Andromeda} employs Sentence-BERT and seasonal-trend-based metric analysis to enable \llm to leverage information from metrics, historical questions, and diagnosis manuals to generate configuration tuning suggestions.}

Although \llm-based methods offer high generalization ability and can generate flexible diagnosis outputs, they have several limitations. 
First, they prompt \llms using some general documents only (e.g., basic \om concepts), based on which \llms (even equipped with advanced techniques like tree of thought~\cite{dbot2024}) easily yield generic results or meet diagnosis failures (e.g., analyzing over non-existent metrics). 
Second, while the LLM+RAG approach allows dynamic document knowledge retrieval~\cite{Andromeda}, typical RAG paradigm conducts top-$k$ matching of the separated knowledge chunks, which destroys the original knowledge relations (e.g., a diagnosis path involving multiple steps) and causes inaccurate or incomplete diagnosis. Besides, similarity-based RAG may return irrelevant knowledge and negatively affect diagnosis (e.g., misleading diagnosis under the guidance of irrelevant ones).





{\it Therefore, we need to develop an experience-enhanced \llm framework that can (1) systematically integrate \om experience without missing the original relations, and (2) support new root causes and recovery solutions for effective and extensible database \om.}

%% file: 3_overview.tex
\section{\oursys\ Overview}
\label{sec:overview}

\begin{figure*}[!t]
  \centering
  \includegraphics[width=\linewidth]{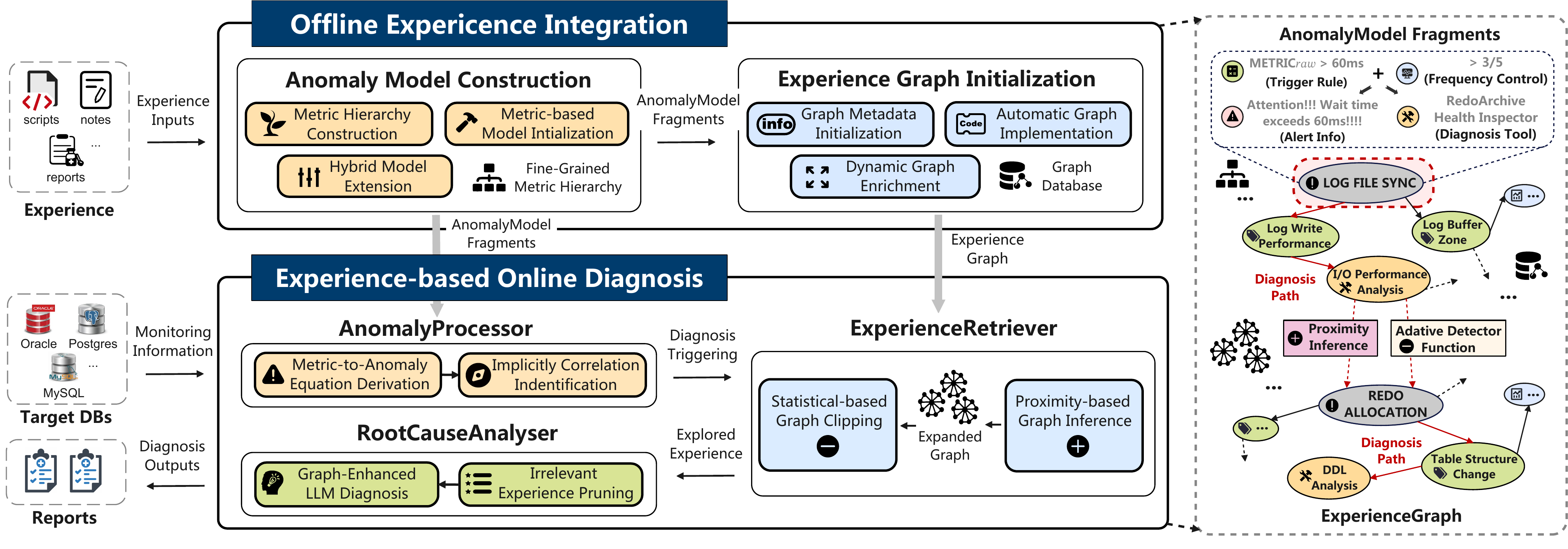}
  \vspace{-2.75em}
  \caption{System Overview of \oursys.}
  \label{fig:overview}
  \vspace{-.75em}
\end{figure*}

\begin{sloppypar}
\hi{Architecture.} \oursys~is composed of five key components (Figure~\ref{fig:overview}). {\ding{182} \experiencegraph encodes expert \om experience into a heterogeneous graph model, where vertices denote \om information (e.g., metrics), and edges capture relations involved in multi-step anomaly analysis;} \ding{183} \anomalymodel performs anomaly detection (using equations derived from metric-anomaly correlation analysis) based on the fine-grained metric hierarchy (e.g., raw data $\rightarrow$ aggregated data) and descriptive anomaly metadata (e.g., symptom illustration);  \ding{184} \anomalyprocessor extracts relevant anomaly analysis information by leveraging both the \anomalymodel outputs and implicitly correlated metrics obtained from standard diagnostic tools (e.g., retrieving \texttt{LOG\_FILE\_SYNC} wait times); 
\ding{185} \experienceretriever\ {automatically explores anomaly analysis paths through a two-stage graph evolution strategy (i.e., proximity-based graph expansion $\rightarrow$ statistical graph clipping) to accumulate relevant experience};
\ding{186} \rootcauseanalyser employs reasoning \llms to simulate DBA-style diagnosis (producing accurate and actionable reports) based on the graph-augmented experience. 
\end{sloppypar}

Note that, with the above components (e.g., multi-metric correlation, graph-based \om experience encoding), \oursys operates effectively using general reasoning \llms~\cite{gpt4o, DeepSeekV3} (see Section~\ref{sec:exp}), eliminating the need for specialized \llm training.

\begin{sloppypar}
\hi{Workflow.} 
As shown in Figure~\ref{fig:overview}, \oursys~operates in two stages: \emph{offline experience integration} and \emph{experience-based online diagnosis}.

\noindent $\blacktriangleright$ \bfit{Offline Stage.}
{Given \om experience from various sources (e.g., resolved anomaly cases), \oursys~first constructs a set of distinct \anomalymodel fragments, capturing different abnormal patterns (e.g., statistical co-occurrence of multiple metrics).
These \anomalymodel fragments are then used to initialize the metadata of the \experiencegraph, which is automatically materialized in a graph database (e.g., via Cypher queries in Neo4j~\cite{Neo4j}).
The graph is further enriched through strategies such as linking vertices with shared properties.}

\noindent $\blacktriangleright$ \bfit{Online Stage.}
Upon receiving a diagnosis request, the \anomalyprocessor performs multi-metric anomaly detection (via the \anomalymodel), mapping the detected anomaly to its corresponding trigger vertex in the \experiencegraph. With abnormal and implicitly correlated metrics at this vertex, the \experienceretriever explores potential diagnostic paths by (1) expanding the graph through proximity-based inference to connect related vertices (e.g., \texttt{REDO\_ALLOCATION} anomaly vertex); and (2) pruning paths containing irrelevant or normal metrics. Using the retrieved metrics and explored paths, the \rootcauseanalyser prompts \llm to (1) extract the most relevant experience from the graph paths, and (2) generate diagnosis reports with root causes and recommended recovery solutions.
\end{sloppypar}

Note that \oursys~can uncover previously unseen root causes and solutions during online diagnosis by (1) traversing diverse graph paths to derive new composite experience, and (2) guiding \llms to infer new root causes through progressive reasoning over similar but non-identical experience (i.e., explored paths) that cannot directly resolve the target anomaly.

%% file: 5_knowledge.tex
\vspace{1em}

\section{Graph Model for \om Experience Characterization}
\label{sec:graph}


A vast amount of \om experience exists in diverse forms, including historical diagnosis reports, resolved anomaly cases, and technical notes~\cite{MOS, DBAExchange, MySQLForum, SQLServerQA, BID, zhou2025cracksql}.
However, a formal and structured representation for systematically characterizing these experiences to enable automated diagnosis is still lacking. To bridge this gap, we propose a heterogeneous graph model (\experiencegraph) that encodes these experience fragments, where anomaly analysis steps are naturally modeled as traversal paths within this graph.



\subsection{Experience Graph Model}
\label{subsec:graph}



\begin{figure}[!t]
  \centering
  \includegraphics[width=.9\linewidth]{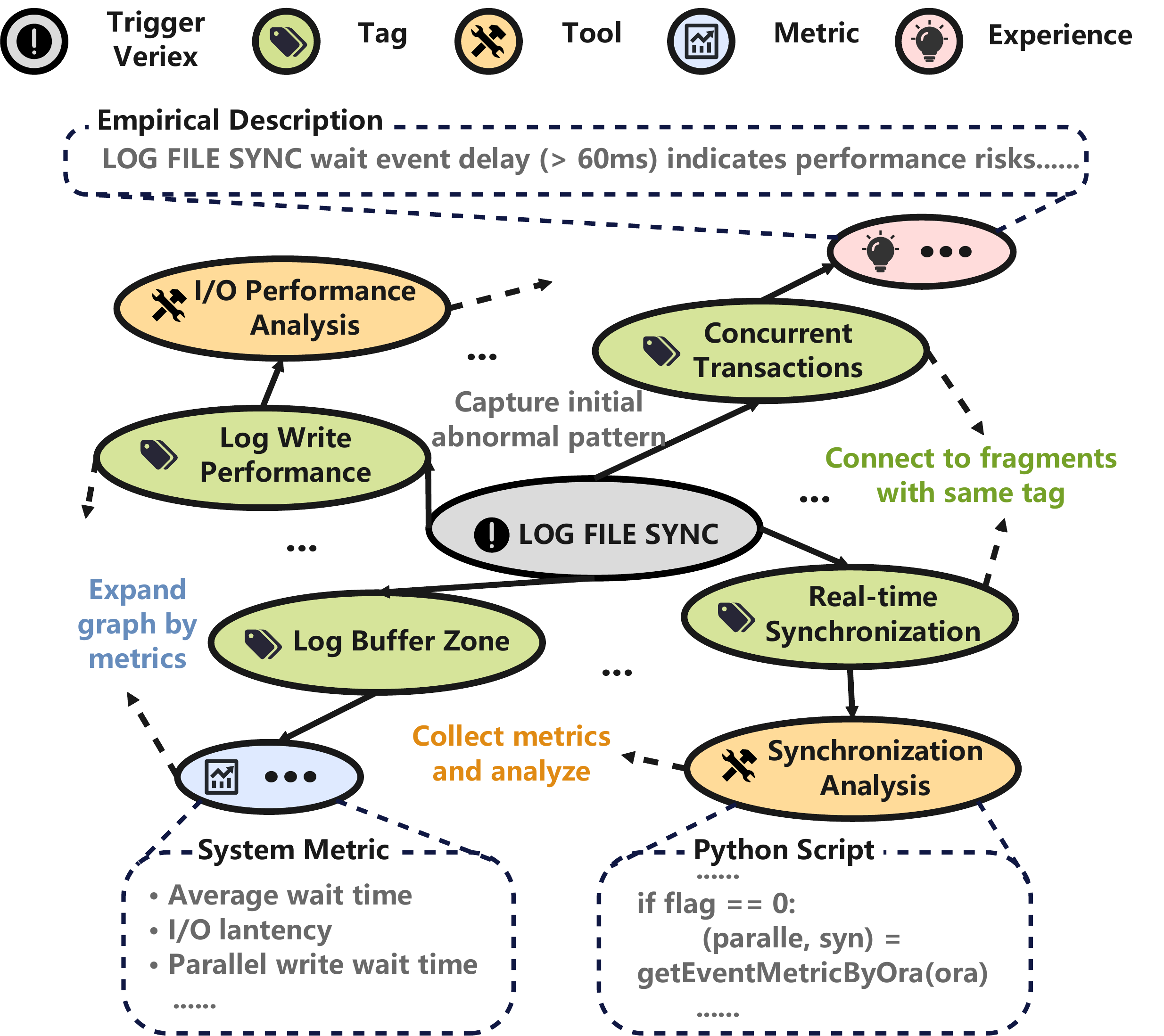}
  \caption{Example \om Graph Model in \oursys.}
  \label{fig:graph_model}
  \vspace{-1.75em}
\end{figure}

\begin{sloppypar}
To integrate \om experience, existing approaches either rely on rules with predefined numeric metric thresholds~\cite{ADDM, DBSherlock} or basic RAG strategies where \llms diagnose by loosely connected document chunks~\cite{Andromeda, panda, gaussmaster}. They fail to capture complex relations that require to consider heterogeneous information in \om experience.  To address this, \oursys~proposes the first \om-specific heterogeneous graph model (\experiencegraph) that can be easily utilized by both \llms and human DBAs. As shown in Figure~\ref{fig:graph_model}, vertices represent essential factors (e.g., abnormal metrics), while edges denote potential diagnosis paths.


Formally, we design \experiencegraph as a directed heterogeneous graph: \(\mathcal{G} = (\mathcal{V}, \mathcal{E}, \mathcal{R}),\) where $\mathcal{V}$ is the set of vertices (in Table~\ref{tab:vertex}), $\mathcal{R}$ is the set of relations, and each directed edge in $\mathcal{E}$ is represented as a triplet 
\(
(v_{\text{src}}, r, v_{\text{tgt}}) \in \mathcal{E} \quad \text{with} \quad r \in \mathcal{R}.
\) 
\end{sloppypar}

\begin{table*}[!t]
\caption{Vertex Types in \om Graph Model.}
\vspace{-.25cm}
\label{tab:vertex}
\resizebox{\linewidth}{!}{
\begin{tabular}{|c|l|l|}
\hline
\rowcolor[HTML]{333333} 
{\color[HTML]{FFFFFF} \textbf{Type}}                                   & \multicolumn{1}{c|}{\cellcolor[HTML]{333333}{\color[HTML]{FFFFFF} \textbf{Content}}}      & \multicolumn{1}{c|}{\cellcolor[HTML]{333333}{\color[HTML]{FFFFFF} \textbf{Example}}}                                                                                                \\ \hline
\textbf{Trigger Vertex}                                                 & Rules or patterns for monitoring database status and analyzing anomaly.                       & \texttt{LOG FILE SYNC} wait delay anomaly.                                                                                                                         \\ \hline
\textbf{Experience}                                               & Information for anomaly analysis, including explanations, solutions, and backgrounds.         & \texttt{LOG FILE SYNC} wait event delay (> 60ms) indicates performance risks.                                                                                \\ \hline
\textbf{Tool}                                                          & Python-based executable tool for specific anomaly diagnosis featuring multiple search paths.  & \texttt{log\_file\_sync\_ana.py} script for high \texttt{LOG FILE SYNC} wait time analysis.                                                       \\ \hline
\textbf{Metric}                                                        & Quantitative data from database objects reflecting operational status or performance.         & \texttt{LOG FILE SYNC} average wait time (e.g., 10ms-60ms).                                                                                                        \\ \hline
\textbf{Tag}                                                           & Tools for identifying and categorizing metrics, knowledge points, enhancing graph connection. & (1) Locks, (2) Concurrency, (3) Hot Block Contention.                                                                                                                               \\ \hline
\textbf{Auxiliary}                                              & Database metric attributes, including collection frequency and associated objects.            & (1) Period Average, (2) Period Median, (3) 90th/95th percentile.                                                                                                                    \\ \hline
\end{tabular}
}
\end{table*}

\noindent \textbf{Vertex Modeling.}
\oursys~currently supports six vertex types. 

\noindent $\bullet$ \bfit{(1) Trigger Vertex} detects potential database anomalies, which captures abnormal metric patterns using hybrid information such as multi-metric equations, triggering frequencies, and textual descriptions. Serving as the entry point for anomaly analysis, it initiates the exploration of different diagnosis paths in the graph. For instance, \texttt{LOG\_FILE\_SYNC} vertex detects slowdowns due to log-writing operations in Oracle databases. Without \emph{Trigger Vertex}, we cannot automatically associate input anomaly with relevant diagnosis paths, making subsequent graph-based reasoning infeasible.

\noindent $\bullet$ \bfit{(2) Metric Vertex} involves statistical indicators to capture database runtime status (e.g., average wait time and I/O latency). {Without \emph{Metric Vertex}, we cannot provide the fine-grained quantitative context needed to interpret \emph{Trigger Vertices} or to support downstream anomaly reasoning and correlation analysis.}

\noindent $\bullet$ \bfit{(3) Experience Vertex} encodes domain-specific \om experience about what the anomaly entails and how to resolve it. For example, the vertex of \texttt{LOG\_FILE\_SYNC} wait event exceeding 60\,ms poses significant performance risks, requiring reducing commit frequency, or adjusting parameters.
{Without \emph{Experience Vertex}, we cannot leverage expert knowledge to guide and enhance the accuracy and completeness of anomaly diagnosis.}

\noindent $\bullet$ \bfit{(4) Tool Vertex} represents executable scripts for collecting and analyzing abnormal metrics. For example, \emph{Synchronization Analysis} tool vertex refers to a Python script that retrieves lock-related wait events and evaluates contention patterns to identify potential synchronization bottlenecks. {Without \emph{Tool Vertex}, we cannot incorporate automated data collection and analysis procedures to perform in-depth metric analysis and validate anomaly hypotheses.}

\noindent $\bullet$ \bfit{(5) Tag Vertex} classifies vertices into semantic categories (e.g., \texttt{Concurrent Transactions}).
It enhances graph connectivity by linking vertices with the same tags, facilitating experience aggregation and cross-case reasoning.
{Without \emph{Tag Vertex}, we cannot exploit category-level correlations among vertices, limiting its ability to generalize knowledge across similar anomalies.}



\noindent $\bullet$ \bfit{(6) Auxiliary Vertex} {provides supplementary information to enrich the interpretation of detected metrics.}
For example, auxiliary vertex of \emph{Metric Attribute} records additional details such as collection frequency and 90th/95th percentile values, offering deeper insight into metric behavior. {Without \emph{Auxiliary Vertex}, we lacks the contextual data necessary for precise metric characterization, which may impair anomaly correlation and the diagnosis accuracy.}



\noindent \textbf{Edge Modeling.}
\oursys~currently supports four edge types.



\begin{sloppypar}
\noindent $\bullet$ \bfit{(1) Containment Edge} represents the inclusion relationship where a \emph{Trigger Vertex} (e.g., the \texttt{LOG\_FILE\_SYNC} anomaly with critical wait timeout) contains related \emph{Experience Vertex} (e.g., guidance about redo log analysis steps). 
\end{sloppypar}



\noindent $\bullet$ \bfit{(2) Relevance Edge} reflects the relation between a \emph{Metric Vertex} (e.g., average wait time) and a \emph{Trigger Vertex}. 

\noindent $\bullet$ \bfit{(3) Diagnosis Edge} defines the relationship where an \emph{Experience Vertex} (e.g., {ash\_db\_io\_ana} for database I/O analysis) utilizes a \emph{Metric Vertex} (e.g., {db file sequential read wait time}) during diagnosis. 

\begin{sloppypar}
\noindent $\bullet$ \bfit{(4) Synonym Edge} represents the semantic equivalence between two \emph{Tag Vertices} expressed differently but referring to the same concept (e.g., physical\_read and disk\_read; shared\_pool and shared\_buffer). {Without \emph{Synonym Edge}, \oursys~cannot unify semantically equivalent tags, resulting in loosely connected graph fragments and limiting the integration of relevant experience.}
\end{sloppypar}

\begin{example}
\begin{sloppypar}
{Figure~\ref{fig:graph_model} demonstrates a simplified graph for \texttt{LOG\_FILE\_SYNC} anomaly in Oracle.
The \emph{Trigger Vertex} captures the detected abnormal metric patterns, while surrounding \emph{Tag Vertices} enhance connectivity by linking them to relevant vertices.
Specifically, the ``\emph{Concurrent Transactions}'' tag vertex associates an \emph{Experience Vertex} describing performance risks from wait delays.
The ``\emph{Log Buffer Zone}'' tag vertex connects to \emph{Metric Vertices} with system statistics (e.g., average wait time) for statistical analysis.
The ``\emph{Real-time Synchronization}'' tag vertex links to a \emph{Tool Vertex} (e.g., \emph{Synchronization Analysis}) with an executable Python script. Overall, \oursys~constructs a graph for Oracle database with over 300,000 edges, including 82 \emph{Trigger Vertices} (covering common anomalies), 550 \emph{Metric Vertices}, 317 \emph{Experience Vertices}, and 897 \emph{Tag Vertices}.}
\end{sloppypar}
\end{example}

Note that \om graph model in \oursys~supports flexible extension or refinement to new types of vertices and edges. For instance, each edge carries one or more attributes to support the addition of new edge types. Besides, the implicitly connected edges can be identified through graph evolution and \llm reasoning (Section~\ref{sec:diagnosis}).





\subsection{Graph Model Construction}
\label{subsec:construction}

Given the vast and complex \om experience pieces, reducing the manual effort required for graph model construction is extremely important~\cite{KGSurvey, LLMKGConstruction1, LLMKGConstruction2}. However, existing methods mainly utilize basic ML (e.g., CauseRank~\cite{CauseRank}) to automatically add graph edges, which rely on simple causal assumptions between vertices and may fail to uncover implicit relations.
{Instead, we propose a semi-automatic graph construction approach, which begins with an initial manual graph sketch of key vertices (e.g., \emph{Trigger Vertices}), while the remaining vertices (e.g., involved \emph{Metric Vertices}) are automatically extracted to expand to a unified graph.}

Specifically, \oursys~{collects the source information including official database documents and historical anomaly case reports (e.g., 15,000 MOS documents for Oracle~\cite{MOS}),} and builds the \om graph model in the following steps.


\noindent$\bullet$ \hi{Graph Metadata Initialization.} Given a set of resolved anomaly cases $\mathcal{A}$\ and their troubleshooting manuals, we first initialize the graph metadata. Specifically, we define \emph{Trigger Vertices} with the rule‑based predicates (e.g., statistical multi-metric relations in Section~\ref{sec:anomaly}) that detect anomalies in $\mathcal{A}$. {Once this vertex is defined, the remaining vertices are automatically extracted and attached based on corresponding information with \emph{Trigger Vertex} as the central point.}  For example, we collect metrics (i.e., \emph{Metric Vertex}) used to detect anomalies and \om experience like {\it ``For the anomaly of excessive REDO generation, we can reduce the amount of REDO''} (i.e., an edge between \emph{Metric Vertex} and \emph{Experience Vertex}).



\noindent$\bullet$ \hi{Graph Implementation.} Next, to efficiently manage and use the graph, we transform the initialized metadata (e.g., vertices, edges, and their corresponding types) into executable statements within a graph database (e.g., Neo4j~\cite{Neo4j}). In this way, we can efficiently implement the graph in a relatively short time. {For instance, building a graph with 150,000 vertices is approximately 15 minutes, which is automated by executing Cypher queries.}

\begin{sloppypar}
Besides, we construct a unified graph for all the 25 supported databases, tagging each vertex with corresponding database. The advantage is that these databases can share the common knowledge ({vertices} with special general tag) and avoid {repeated construction}.
\end{sloppypar}



\noindent$\bullet$ \hi{Graph Enrichment.} Building on the graph, we automatically enrich it by adding additional edges, such as linking vertices with shared \emph{Tag Vertices} or connecting \emph{Metric Vertices} with highly similar statistical profiles. For instance, we can derive over 300,000 edges from just dozens of anomaly models in Oracle.

\noindent$\bullet$ \hi{Graph Update.}
During online usage, \oursys~support incremental updates {by incorporating newly acquired experience fragments, which are automatically linked to existing vertices via the automatic graph enrichment mechanism above.
Furthermore, existing vertices can be updated or refined via graph database operations (i.e., Cypher queries).
For instance, enriching the knowledge description of \emph{Experience Vertex} or removing outdated \emph{Metric Vertex}.}

In this way, we construct a graph model that {incorporates DBA expertise in managing 5,000+ databases over the past 10 years, along with 2,000+ historical anomaly cases and supports new anomalies via a graph evolution mechanism (see Section~\ref{subsec:evolution}). For instance, it currently contains 2,911 vertices (for Oracle), 1,786 (for MySQL), 2,272 (for PostgreSQL), and 2,259 (for DM8).}

In practice, the well-built graph can be easily utilized by both human DBAs and \llms, who can explore the potential anomaly analysis paths over the graph (e.g., {\it anomaly (Trigger Vertex) $\rightarrow$ category (Tag Vertex) $\rightarrow$ experience $\cdots$ $\rightarrow$ metrics $\cdots$}) and significantly reduce the \om overhead.

%% file: 4_observability.tex
\section{Correlation-Aware Anomaly Model}
\label{sec:anomaly}


We next introduce \anomalymodel embedded in the {\it Trigger Vertex} of \om graph.
The model aims to address the critical metric-anomaly correlation challenge. Unlike most existing methods that focus solely on detecting metrics with abnormal values~\cite{ADDM, DBSherlock}, our model captures the relationships among metrics to uncover anomalies that emerge from correlated behaviors (e.g., simultaneous spikes in \emph{log file sync delay} and \emph{log file parallel write} to reveal systemic issues like I/O bottlenecks).







\subsection{Multi-Metric Anomaly Detection}
\label{subsec:multi_metric}

Metrics serve as primary factor to facilitate effective database \om.
However, there exists a high volume of metrics from diverse monitoring sources, and these metrics need to be further processed to derive essential information (e.g., trend changes).
To address this issue, \oursys~first constructs a unified metric hierarchy (encompassing system metrics, logs, and traces), and then performs statistical multi-metric correlation analysis to automatically derive effective anomaly detection equations.







\hi{Fine-Grained Metric Hierarchy.} 
To provide a comprehensive view of database status, \oursys~collects and processes metrics from multiple sources (e.g., logged events and execution statistics) and introduces a carefully designed metric hierarchy. 
Specifically, \oursys~organizes database metrics into increasingly fine-grained subcategories, with main classes at the top level (e.g., configuration-related).
This hierarchical metric organization simplifies {large-scale metric handling by aligning only metrics in the hierarchical tree with relevant categories to the detected anomaly, reducing the noise of irrelevant metrics and improving diagnosis efficiency}. 

These metrics are initially retrieved as \emph{raw data} from external tools (e.g., Prometheus~\cite{prometheus}), retaining only essential details like category IDs and error messages.
Subsequently, additional statistical data (e.g., incremental differences, rolling averages, and histograms) are computed lazily, i.e., \emph{only} when needed for diagnosis.


\begin{itemize}[leftmargin=10pt,topsep=2pt]
\item \emph{(1) Immediate Raw Data Collection:}
Data such as execution statistics, log records, and workload traces are gathered directly and condensed to store only relevant information.

\item \emph{(2) Lazy Statistical Data Calculation:}
Periodic, higher-level metrics (e.g., incremental deltas and averages) are generated on demand, avoiding unnecessary overhead. For example, \oursys~collects long-interval AWR data (e.g., 30-minute intervals) and short-interval ASH data (e.g., less than 10 seconds) for Oracle databases.
\end{itemize}




\begin{sloppypar}
\hi{Metric-to-Anomaly Correlation.} To effectively capture metric-to-anomaly relation, \oursys~develops a collection of anomaly models. Each model captures a specific database anomaly based on distinct multi-metric patterns or longitudinal single-metric comparisons. Unlike typical threshold-based methods that detect excessive anomalies with limited accuracy, the anomaly models in \oursys\ (1) leverage both established \om experience and analysis over multiple metrics, and (2) are automatically generated from basic elements (e.g., the \texttt{LOG\_FILE\_SYNC} equation in Figure~\ref{fig:overview}). 
\end{sloppypar}

\noindent{\bfit{(1) The Derivation of Anomaly Detection Functions.}} 
\oursys defines anomaly detection equations with configurable parameters, triggering a specific anomaly model only when these expressions are evaluated as \texttt{TRUE}. 
The equations incorporate items such as {\it system metric values, configuration settings, and statistical functions} within a time interval. For instance, the equation for \texttt{LOG\_SYNC\_FILE} is written below:
\begin{equation*}
\begin{aligned}
Is\_Anomaly(\texttt{LOG\_SYNC\_FILE}) =  
& \Big( \mathbf{METRIC}_{raw} > \mathit{time\_threshold}_1 \Big) 
\end{aligned}
\end{equation*}
\vspace{-1em}
\begin{equation*}
\begin{aligned}
& \lor \Big[ \big( \mathbf{METRIC}_{10min} = \mathit{trend} \big) 
\land \big( \mathbf{METRIC}_{raw} > \mathit{time\_threshold}_2 \big) \Big],
\end{aligned}
\end{equation*}
\[
\begin{aligned}
{{trend}} \in \big\{ & \textbf{0}\,{{(stable)}}, \; \textbf{1}\,{{(sharp decline)}}, \; \textbf{2}\,{{(slow decline)}}, \\
                     & \textbf{3}\,{{(sharp rise)}}, \; \textbf{4}\,{{(slow rise)}}, \; \textbf{5}\,{{(fluctuating)}} \big\}
\end{aligned}
\]

\noindent where $\mathbf{METRIC}_{raw}$ denotes the collected raw data for \textit{log file sync average wait time} and $\mathbf{METRIC}_{10min}$ denotes the composite data derived from raw data through volatility analysis algorithms~\cite{volatility}. 
{Different thresholds are adaptively and automatically adjusted across scenarios, which identifies abnormal metrics based on statistical patterns (see Section~\ref{subsec:evolution}).} It quantifies the patterns of fluctuation or trend. The anomaly is detected under two cases: (1) the average wait time is longer than $time\_threshold\_1$ (60ms); (2) the composite metric showcases a sharp rise trend ($trend = 3$) in 10 minutes, and the average wait time is longer than $time\_threshold\_2$ (6ms).

\noindent{\bfit{(2) Frequency Control.}} 
\oursys~employs frequency control to evaluate whether the metric value holds in multiple assessments and reduce the volume of false or irrelevant anomaly detections. For instance, the above detection equation is raised only if the condition holds in 3 out of 5 consecutive evaluations by configuring the settings of trigger frequency to 3/5.

\begin{table}[!t]
\caption{Statistics of anomaly models in \oursys. Variations in numbers arise from differences in available resources.}
\vspace{-1em}
\label{tab:anomaly_stats}
\resizebox{\linewidth}{!}{
\begin{tabular}{|c|c|c|c|c|c|c|c|}
\hline
\rowcolor[HTML]{333333} 
{\color[HTML]{FFFFFF} \textbf{Database}}                          & {\color[HTML]{FFFFFF} \textbf{Oracle}} & {\color[HTML]{FFFFFF} \textbf{DB2}} & {\color[HTML]{FFFFFF} \textbf{SQL Server}} & {\color[HTML]{FFFFFF} \textbf{MySQL}} & {\color[HTML]{FFFFFF} \textbf{PostgreSQL}} & {\color[HTML]{FFFFFF} \textbf{OceanBase}} & {\color[HTML]{FFFFFF} \textbf{GaussDB}} \\ \hline
\textbf{Metric}                                                   & 550                                    & 927                                 & 314                                        & 316                                   & 645                                        & 963                                       & 658                                     \\ \hline
\textbf{\begin{tabular}[c]{@{}c@{}}Diagnosis\\ Tool\end{tabular}} & 396                                    & 10                                  & 124                                        & 215                                   & 148                                        & 98                                        & 151                                     \\ \hline
\textbf{\begin{tabular}[c]{@{}c@{}}Anomaly\\ Model\end{tabular}}  & 82                                     & 7                                   & 25                                         & 91                                    & 36                                         & 34                                        & 85                                      \\ \hline
\end{tabular}
}
\vspace{-1em}
\end{table}

\subsection{Implicitly-Correlated Metric Identification}
\label{subsec:implicit_metric}


To further identify implicitly correlated metrics apart from the metrics involved in the anomaly detection equations, \oursys~integrates a series of executable low-code tools (diagnosis insights of specific metric sets) to obtain useful diagnosis information, i.e., obtaining relevant metrics and linking the relevant diagnosis insights in \emph{Experience Vertices}.

Specifically, these tools leverage DSL programming to define complex abnormal scenarios. Existing scripting languages (e.g., Python) are utilized with standardized input-output formats, relying on monitoring platforms for metric and configuration data access. For example, for \texttt{LOG\_FILE\_SYNC}, \oursys includes two tools to enhance diagnosis accuracy and efficiency.

\begin{itemize}[leftmargin=10pt,topsep=2pt]
    \item \emph{(1) LogSync Performance Verifier:} The tool automatically detects and analyzes database performance bottlenecks by monitoring key metrics (e.g., log file sync wait time, redo generation rate, transaction commits) against baseline threshold, identifying anomalies, and providing root cause insights (e.g., excessive commits, redo overload, undersized log buffers);
    \item \emph{(2) RedoArchive Health Inspector:} The tool analyzes discrepancies between archive log and redo log sizes, evaluates parameter configurations (e.g., \texttt{log\_buffer}, \texttt{archive\_lag\_target}), and identifies performance risks caused by \texttt{rapid\_redo\_log} switching or suboptimal settings, while providing technical rationales and compliance checks.
    The analysis includes verification of log file size and detection of any abnormal log patterns.
\end{itemize}

\begin{sloppypar}
{\it \bfit{Example.}
Consider the \texttt{LOG SYNC FILE} anomaly model for Oracle databases, which detects slowdowns caused by log-writing operations. When a transaction commits or rolls back, forcing a session to wait for the log writer to flush redo logs, the system can degrade significantly if I/O capacity is insufficient. The associated anomaly equation states that if the immediate wait time exceeds 60\,ms, or if a volatility analysis indicates a sharp 10-minute rise and the current wait time surpasses 6\,ms, an alert should be triggered.
To avoid transient fluctuations, \oursys~raises an alert only if this condition holds in 3 out of 5 consecutive evaluations. Subsequently, diagnosis tools (e.g., a \emph{LogSync Performance Diagnosis}) analyze relevant parameters (e.g., redo generation rate) and configurations (e.g., \texttt{archive\_lag\_target}) to uncover the root cause and recommend solutions.}
\end{sloppypar}

As shown in Table~\ref{tab:anomaly_stats}, \oursys~currently includes more than 800 anomaly models that cover the common anomalies in main-stream databases (e.g., over 70 anomaly models for Oracle). Note \oursys only requires basic tools for metric collection, much fewer than traditional rule-based methods.

%% file: diagnosis.tex
\section{Scenario-Aware Anomaly Diagnosis}
\label{sec:diagnosis}

\begin{figure}[!t]
  \centering
  \includegraphics[width=\linewidth]{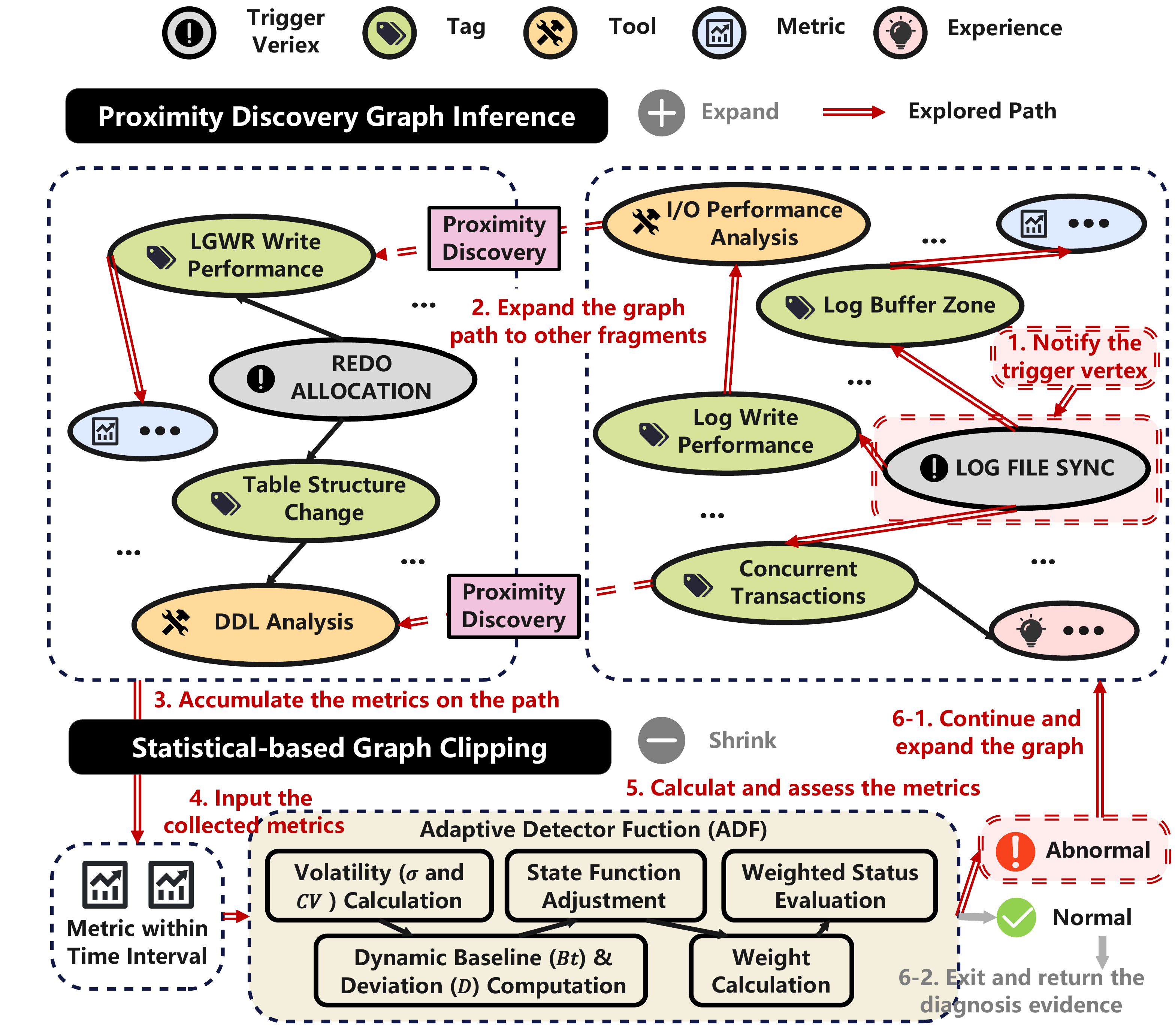}
  \caption{Two-Stage Graph Evolution in \oursys.}
  \label{fig:graph_evolution}
\end{figure}

The next problem is how to automatically perform accurate diagnosis using the above graph model (and anomaly analysis methods). There are two main challenges. First, anomalies in real-world systems are often interrelated, issues in one anomaly model (represented as a connected subgraph, such as \texttt{LOG\_FILE\_SYNC}) may trigger or exacerbate issues in others (e.g., \texttt{REDO\_ALLOCATION}), making cross-subgraph reasoning complex. Second, the graph model may produce false positives (e.g., irrelevant metrics) or incomplete results, and its outputs typically require high expertise to interpret.

\subsection{Two-Stage Graph Evolution}
\label{subsec:evolution}


\begin{sloppypar}
In real-world scenarios, anomalies are rarely isolated, where a performance issue in one anomaly model may simultaneously trigger or worsen issues in another.
However, different anomaly models such as those related to \texttt{LOG\_FILE\_SYNC} and \texttt{REDO\_ALLOCATION} may appear only loosely connected in the initialized graph, which share sparse and fragmented experience (e.g., concurrency-related wait events). To this end, we propose an automatic ``graph evolution'' mechanism to \emph{dynamically discover and connect} such related experience fragments across different anomaly models. This mechanism comprises two main stages.
\end{sloppypar}


\begin{sloppypar}
\noindent
\underline{(1)~\emph{Graph Inference and Proximity Discovery.}} 
Using specialized graph query language (Cypher), \oursys~collects and aggregates relevant metrics (e.g., I/O wait times and commit latency) by traversing related nodes and edges based on configurable thresholds, integrating the results into a unified metric view.
When overlapping or related anomaly scenarios are identified (e.g., an I/O-related concurrency issue that also appears in the \texttt{REDO ALLOCATION} model), the system creates or reinforces cross-edges between the associated experience fragments.
Through repeated iterations, the \om graph model evolves into a denser and more interconnected structure. For instance, {when diagnosing a \texttt{LOG\_FILE\_SYNC} anomaly, the system leverages a dense graph where vertices such as \texttt{log file parallel write}, \texttt{Mem unavailable}, and \texttt{commit latency} are tightly connected through historical causality. This enriched structure enables accurate identification of combined root causes such as I/O bottlenecks and memory pressure, which may be overlooked in sparser graphs.}


\noindent
\underline{(2)~\emph{Adaptive Abnormal Metric Detection.}} 
After metrics over the evolved graph paths are obtained, the \emph{Adaptive Detector Function} (ADF) identifies which metrics exhibit abnormal behaviors and determines whether further graph expansion is warranted. Specifically, given a metric value sequence \( X = \{x_1, x_2, \ldots, x_n\} \) observed over time intervals \( t = [t_1, \dots, t_n]. \), the ADF follows the following steps.
\end{sloppypar}

\noindent
\textbf{Step 1. Volatility (\( \sigma \) and \( C_V \)) Calculation.} 
We first computes the standard deviation of \(X\) (as \(\sigma\)) to quantify its fluctuation amplitude:
\[
  \sigma = \sqrt{\frac{1}{n-1} \sum_{i=1}^n (x_i - \bar{x})^2},
\]
where \(\bar{x}\) is the mean value. It also obtains \(C_V = \rho_V / \rho_R\), where \(\rho_V\) is the autocorrelation coefficient of the \emph{volatility} series, and \(\rho_R\) is the autocorrelation coefficient of a random volatility series. It helps quantify how persistent or random the fluctuations are over time.

\noindent
\textbf{Step 2. Dynamic Baseline (\( B_t \)) and Deviation (\( D \)) Computation.}
\(\oursys\) derives an adjusted dynamic baseline \(B_t\) for each time interval \(t\). This baseline is updated hourly across different databases to maintain adaptiveness. The deviation is then calculated as \(D = |\,x_t - B_t|\). By design, \(B_t\) can incorporate parameterized factors that capture known operational patterns (e.g., higher \texttt{log file sync} overhead during batch jobs).

\noindent
\textbf{Step 3. State Function Adjustment.}
A state function \(F_{\text{state}}(x_t, B_t)\) classifies how close \(x_t\) is to the baseline:
\[
  F_{\text{state}}(x_t, B_t) = 
  \begin{cases} 
    1 - \tfrac{D}{\sigma}, & \text{if \(x_t\) is near \(B_t\)}; \\[5pt]
    \tfrac{D}{\sigma}, & \text{otherwise}.
  \end{cases}
\]
Smaller deviations from \(B_t\) yield a larger value under the first case; conversely, large divergences indicate potential anomalies.

\noindent
\textbf{Step 4. Weight Calculation.}
The volatility weight \(w_1\) is computed dynamically based on \(\sigma\) and a threshold \(\theta\):
\[
 w_1 = \frac{\sigma}{\sigma + \theta}, \quad w_2 = 1 - w_1.
\]
When \(\sigma > \theta\), \(\oursys\) assigns more weight to volatility, signifying that larger fluctuations in the metric are more relevant.

\noindent
\textbf{Step 5. Weighted Status Evaluation.}
Finally, an overall anomaly score \(S\) is computed:
\[
 S = w_1 \cdot \sigma \;+\; w_2 \cdot F_{\text{state}}(x_t, B_t).
\]
If \(S\) exceeds a threshold (tunable per environment), the metric is marked ``abnormal'', prompting further graph expansion (e.g., adding or revisiting neighboring nodes for additional checks). If no metric is flagged, evolution terminates.

\begin{sloppypar}
{\it \bfit{Example.}
Suppose \oursys~initially detects a suspicious \texttt{LOG\_FILE\_SYNC} seed node during \textbf{Graph Inference and Proximity Discovery}.
By querying via Cypher and traversing relevant edges (e.g., ``\texttt{avg\_log\_sync\_time}'' and ``\texttt{txn\_throughput}''), the system finds a \texttt{REDO\_ALLOCATION} anomaly model sharing a concurrency wait event. Recognizing this overlap, \oursys~consolidates their metrics (e.g., ``\texttt{redo\_buffer\_busy}'') and establishes a cross-edge between \texttt{LOG\_FILE\_SYNC} and \texttt{REDO\_ALLOCATION}. Next, in \textbf{Adaptive Abnormal Metric Detection}, assume \texttt{avg\_log\_sync\_time} has values $X = \{12,\, 14,\, 55,\, 58,\, 61\}$. The mean is $\bar{x}=40$, and the standard deviation $\sigma\approx21$. With a dynamic baseline $B_t\approx15$, the deviation $D=|\,58-15|\approx43$ is significantly large. Because $\sigma>\theta$ (assume $\theta=10$), the anomaly score $S$ surpasses the threshold. \oursys~then expands the graph, automatically linking new \texttt{REDO\_ALLOCATION} nodes and a diagnostic script (redo\_allocator\_check). Over time, such expansions yield a richer MKG, allowing future queries to reuse these newly formed connections.}
\end{sloppypar}

In this way, we can traverse the graph to extract all relevant information for diagnosis. For instance, retrieving all associated wait events, recommended diagnosis scripts, and concurrency settings required to investigate a spike in \texttt{LOG\_FILE\_SYNC} time. 



%% file: 6_reasoning.tex
\subsection{Graph-Augmented LLM Diagnosis}
\label{subsec:reasoning}

With the explored graph paths, several challenges remain for accurate anomaly diagnosis. First, there may be false positives, such as vertices that appear relevant but do not accurately reflect the root cause. Second, experience within those vertices can be incomplete or difficult for general users to interpret. To this end, \oursys~proposes a prompt-based strategy that guides the reasoning \llm to analyze the experience paths and generate clear, actionable diagnosis reports that include both the identified root causes and corresponding recovery solutions.

To address false positives and incomplete coverage, \oursys~provides \llm with (1) extensive textual analysis experience collected during graph traversal and (2) a collection of accumulated metrics and execution details (e.g., logs, historical performance baselines).
When \llm crafts a diagnosis report, it not only refers the \emph{Triggered Vertex} in the graph but also traces relevant edges to other anomalies.
It then contextualizes these findings by describing how each anomaly interacts within the broader environment (e.g., ``\emph{The concurrency waits grew after I/O latencies exceeded 30\,ms, indicating shared resource contention.}''). This synergy between structured graph data and open-ended generative reasoning allows \oursys~to produce more thorough and comprehensible diagnosis.

\begin{table*}[!t]
\caption{Common Root Causes Observed in Real-World Usage Across the Four Database Systems.}


\label{tab:anomaly_exp}
\resizebox{\linewidth}{!}{
\begin{tabular}{|c|c|c|c|c|c|c|c|c|c|c|c|c|c|c|c|c|c|}
\hline
Database &
  \begin{tabular}[c]{@{}c@{}}HIGH \\ DATA \\ SELECT\end{tabular} &
  \begin{tabular}[c]{@{}c@{}}LOW REDO \\ FILE \\ SIZE\end{tabular} &
  \begin{tabular}[c]{@{}c@{}}LOW REDO \\ GROUP \\ COUNT\end{tabular} &
  \begin{tabular}[c]{@{}c@{}}LOG BUFFER \\ SETTING NOT \\ ENOUGH\end{tabular} &
  \begin{tabular}[c]{@{}c@{}}TABLE \\ INITTRANS \\ NOT ENOUGH\end{tabular} &
  \begin{tabular}[c]{@{}c@{}}BUFFER \\ BUSY \\ WAIT\end{tabular} &
  \begin{tabular}[c]{@{}c@{}}ENQ \\ LOCK \\ WAIT\end{tabular} &
  \begin{tabular}[c]{@{}c@{}}LATCH \\ WAIT\end{tabular} &
  \begin{tabular}[c]{@{}c@{}}HIGH \\ MEMORY\\ USAGE\end{tabular} &
  \begin{tabular}[c]{@{}c@{}}HIGH \\ CPU \\ USAGE\end{tabular} &
  \begin{tabular}[c]{@{}c@{}}BGWRITER \\ PARAMETER \\ PROBLEM\end{tabular} &
  \begin{tabular}[c]{@{}c@{}}SHARED \\ BUFFER \\ NOT ENGHOU\end{tabular} &
  \begin{tabular}[c]{@{}c@{}}CHECKPOINT \\ PARAMETER \\ PROBLEM\end{tabular} &
  \begin{tabular}[c]{@{}c@{}}WAL \\ PARAMETER \\ PROBLEM\end{tabular} &
  \begin{tabular}[c]{@{}c@{}}TABLE \\ DEAD \\ TUPLE\end{tabular} &
  \begin{tabular}[c]{@{}c@{}}INDEX \\ PROBLEM\end{tabular} &
  \begin{tabular}[c]{@{}c@{}}STATISTICS \\ EXPIRED\end{tabular} \\ \hline
Oracle     &
  $\checkmark$ & $\checkmark$ & $\checkmark$ & $\checkmark$ &
  $\checkmark$ & $\checkmark$ & $\checkmark$ & $\checkmark$ &
  $\checkmark$ & $\checkmark$ &
  $\times$ & $\times$ & $\times$ & $\times$ & $\times$ & $\times$ & $\times$ \\ \hline
DM8         &
  $\checkmark$ & $\times$      & $\times$      & $\checkmark$ &
  $\times$      & $\checkmark$ & $\checkmark$ & $\checkmark$ &
  $\checkmark$  & $\checkmark$  &
  $\times$ & $\times$ & $\times$ & $\times$ & $\times$ & $\times$ & $\times$ \\ \hline
Mysql      &
  $\checkmark$ & $\times$      & $\times$      & $\times$      &
  $\times$      & $\checkmark$ & $\checkmark$ & $\checkmark$ &
  $\checkmark$  & $\checkmark$  &
  $\times$ & $\times$ & $\times$ & $\times$ & $\times$ & $\times$ & $\times$ \\ \hline
PostgreSQL &
  $\times$     & $\times$      & $\times$      & $\times$      &
  $\times$      & $\times$     & $\times$      & $\times$      &
  $\times$      & $\times$      &
  $\checkmark$ & $\checkmark$ & $\checkmark$ & $\checkmark$ & $\checkmark$ & $\checkmark$ & $\checkmark$ \\ \hline
\end{tabular}}
\end{table*}

\noindent
\textbf{Prompting \llm for Structured Report Generation.}
A core design feature of \oursys~is the structured prompts to guide the \llm~in generating diagnosis reports that are both actionable and easy to understand, which are composed of the following components: 
Given an observed anomaly, we concatenate five necessary components into a $\bfit{prompt}=\langle\mathcal{S}^a,\mathcal{S}^l,\mathcal{S}^m,\mathcal{S}^e,\mathcal{S}^o\rangle$, where:

\noindent $\bullet$ $\mathcal{S}^a$ (\underline{Anomaly}) specifies the symptom descriptions (e.g., ``CPU usage spiked to 95\% at 16{:}00 on 2023‑10‑05'');  

\noindent $\bullet$ $\mathcal{S}^l$ (\underline{Condition}) encodes the anomaly detection condition (e.g., ``exceeds 90\% for >5 min'');

\noindent $\bullet$ $\mathcal{S}^m$ (\underline{Metrics}) records key statistics, e.g., metric name (\emph{CPU Usage, \%}), time range (\emph{1684600070–1684603670}), and threshold (\emph{90\%});  

\noindent $\bullet$ $\mathcal{S}^e$ (\underline{Experience}) provides contextual facts such as normal load (10 k req/min) and recent maintenance (kernel update 2023‑10‑04);

\noindent $\bullet$ $\mathcal{S}^o$ (\underline{Output}) prescribes the desired report \emph{components}. 

We supply the $\bfit{prompt}$ for \llm to generate diagnosis reports that includes the following contents. 
(1) Anomaly Validation: determines whether the reported anomaly requires further investigation; 
(2) Root Cause Analysis: identifying up to five likely causes supported by metrics, logs, or known fault signatures; 
(3) Recover Solution: suggesting technical adjustments such as configuration changes or query optimizations; 
(4) Summary: providing a concise assessment of overall system health; 
(5) SQL Context: including relevant SQL statements or execution plans if the issue involves database operations.

{\it \bfit{Example.} Consider a scenario in which CPU usage surges to 95\% for over five minutes, coinciding with an abrupt spike in \texttt{LOG\_FILE\_SYNC} wait events. \oursys~collects this numerical data along with experience fragments describing typical concurrency issues under high CPU loads, which is used to prompt \llm to consult the \om graph to check for known concurrency conflicts in situations where CPU usage is near saturation. By examining historical usage patterns, system logs, and wait-event correlations, the \llm concludes that the CPU spike led to excessive wait time for log writes. It then synthesizes a concise {root cause explanation} (e.g., ``{High CPU usage limited log writer throughput, causing queueing in the log buffer}'') and provides an actionable recommendation (``{Scale out the CPU or stagger heavy write workloads to avoid saturating the log writer process}''). Crucially, such reasoning goes beyond a single rule or a static decision tree, instead leveraging {long-term} evidence trails and domain knowledge from the \om graph to construct a thorough diagnosis narrative.}

%% file: 7_experiment.tex
\section{Experiments}
\label{sec:exp}

\subsection{Experiment Setup}
\label{subsec:anomaly}

\begin{sloppypar}
\hi{Databases.} We test four database systems (i.e., Oracle~\cite{Oracle}, MySQL~\cite{MySQL}, PostgreSQL~\cite{PostgreSQL}, and DM8~\cite{Dameng}). The metrics and logs are collected by {adapted tools like Prometheus~\cite{prometheus}}.
\end{sloppypar}

\hi{Anomalies.} Table~\ref{tab:anomaly_exp} lists the detailed root causes of the experimented anomalies across four database systems. The total number of tested scenarios are 178, 114, 127, and 139 for Oracle, MySQL, PostgreSQL, and DM8, respectively. These anomalies can be classified into five categories.

\noindent \textbf{\ding{182} Log Synchronization and Management Issues.} This category covers performance bottlenecks in log writing, synchronization, and management, including: (1) log sync delays causing commit/rollback waits for LGWR (in Oracle) to write redo logs; (2) excessive active log groups in Oracle with unusually high counts of ``ACTIVE'' redo logs; and (3) abnormal REDO log growth (e.g., DM8) due to excessive log generation.



\noindent \textbf{\ding{183} Resource Contention and Concurrency Issues.} These issues arise when multiple sessions or processes compete for shared resources, often leading to lock contention or long waits that reduce concurrency and slow response times. Cases include: 
(1) Oracle hot block contention, where many sessions repeatedly access the same data block; (2) Sudden spikes in active sessions in Oracle, DM8, and MySQL, which can overwhelm the system and severely degrade performance under heavy workloads.


\noindent \textbf{\ding{184} SQL Optimization Issues.} This category involves performance degradation from poorly designed SQL queries or inefficient execution plans. Cases include: (1) Abnormal logical reads in Oracle, where queries fetch far more data blocks from the buffer cache than necessary; and  (2) PostgreSQL full table scans, where queries scan entire tables instead of using indexes, causing unnecessary resource load.


\noindent \textbf{\ding{185} Hardware and System Resource Bottlenecks.} Database performance can be limited by hardware or OS resource constraints, especially under peak load. For instance, for abnormal CPU spikes in MySQL, the sudden surges in processor usage suggest capacity issues or inefficient resource allocation requiring prompt action.


\noindent \textbf{\ding{186} Database Write Performance Issues.} These issues arise from inefficiencies in write operations, slowing transaction commits and reducing system responsiveness. A case is PostgreSQL’s excessive dirty page writes, where backend processes frequently flush modified pages to disk due to insufficient background writer or checkpointer activity, causing latency and lowering throughput as foreground tasks are interrupted.

We ensure the tested anomalies are distinct from those in the graph model; specifically, the graph does not explicitly contain identical root causes or solutions as the test cases. The ground truth results are derived from the \om reports authored by expert DBAs.

\begin{table*}[!t]
\caption{Overall Diagnosis Performance of Different Methods over Anomalies across Four Database Systems {(N/A denotes that diagnosis over the database is not supported by the corresponding method, e.g., D-Bot~\cite{dbot2024} only supports PostgreSQL)}.}
\label{tab:overall}
\resizebox{\linewidth}{!}{
\begin{tabular}{|cc|cccc|cccc|cccc|cccc|}
\hline
\multicolumn{2}{|c|}{\multirow{2}{*}{\textbf{Method}}}                                                                                                                                    & \multicolumn{4}{c|}{\textbf{Oracle}}                                                                                                       & \multicolumn{4}{c|}{\textbf{MySQL}}                                                                                                        & \multicolumn{4}{c|}{\textbf{PostgreSQL}}                                                                                                   & \multicolumn{4}{c|}{\textbf{DM8}}                                                                                                          \\ \cline{3-18} 
\multicolumn{2}{|c|}{}                                                                                                                                                                    & \multicolumn{1}{c|}{\textbf{Precision}} & \multicolumn{1}{c|}{\textbf{F1-Score}} & \multicolumn{1}{c|}{\textbf{Accuracy}} & \textbf{HEval} & \multicolumn{1}{c|}{\textbf{Precision}} & \multicolumn{1}{c|}{\textbf{F1-Score}} & \multicolumn{1}{c|}{\textbf{Accuracy}} & \textbf{HEval} & \multicolumn{1}{c|}{\textbf{Precision}} & \multicolumn{1}{c|}{\textbf{F1-Score}} & \multicolumn{1}{c|}{\textbf{Accuracy}} & \textbf{HEval} & \multicolumn{1}{c|}{\textbf{Precision}} & \multicolumn{1}{c|}{\textbf{F1-Score}} & \multicolumn{1}{c|}{\textbf{Accuracy}} & \textbf{HEval} \\ \hline
\multicolumn{1}{|c|}{\textbf{Traditional}}                                                                  & \textbf{\begin{tabular}[c]{@{}c@{}}Rule-based\\ Tool + DBA\end{tabular}}    & \multicolumn{1}{c|}{0.88}               & \multicolumn{1}{c|}{0.89}              & \multicolumn{1}{c|}{0.88}              & 0.88           & \multicolumn{1}{c|}{1.00}               & \multicolumn{1}{c|}{0.67}              & \multicolumn{1}{c|}{1.00}              & 0.50           & \multicolumn{1}{c|}{1.00}               & \multicolumn{1}{c|}{1.00}              & \multicolumn{1}{c|}{1.00}              & 0.95           & \multicolumn{1}{c|}{1.00}               & \multicolumn{1}{c|}{1.00}              & \multicolumn{1}{c|}{1.00}              & 0.90           \\ \hline
\multicolumn{1}{|c|}{\multirow{2}{*}{\textbf{LLM Only}}}                                                    & \textbf{DepSeek-R1 32B}                                                     & \multicolumn{1}{c|}{0.68}               & \multicolumn{1}{c|}{0.70}              & \multicolumn{1}{c|}{0.65}              & 0.52           & \multicolumn{1}{c|}{0.84}               & \multicolumn{1}{c|}{0.91}              & \multicolumn{1}{c|}{0.71}              & 0.85           & \multicolumn{1}{c|}{0.10}               & \multicolumn{1}{c|}{0.13}              & \multicolumn{1}{c|}{0.83}              & 0.05           & \multicolumn{1}{c|}{0.74}               & \multicolumn{1}{c|}{0.72}              & \multicolumn{1}{c|}{0.01}              & 0.63           \\ \cline{2-18} 
\multicolumn{1}{|c|}{}                                                                                      & \textbf{DeepSeek-R1 671B}                                                   & \multicolumn{1}{c|}{0.77}               & \multicolumn{1}{c|}{0.83}              & \multicolumn{1}{c|}{0.75}              & 0.78           & \multicolumn{1}{c|}{0.67}               & \multicolumn{1}{c|}{0.80}              & \multicolumn{1}{c|}{0.56}              & 0.70           & \multicolumn{1}{c|}{0.75}               & \multicolumn{1}{c|}{0.86}              & \multicolumn{1}{c|}{0.63}              & 0.75           & \multicolumn{1}{c|}{0.60}               & \multicolumn{1}{c|}{0.60}              & \multicolumn{1}{c|}{0.73}              & 0.45           \\ \hline
\multicolumn{1}{|c|}{\textbf{\begin{tabular}[c]{@{}c@{}}LLM\\ (RAG-based)\end{tabular}}}                    & \textbf{ChatDBA}                                                            & \multicolumn{1}{c|}{N/A}                & \multicolumn{1}{c|}{N/A}               & \multicolumn{1}{c|}{N/A}               & N/A            & \multicolumn{1}{c|}{0.50}               & \multicolumn{1}{c|}{0.60}              & \multicolumn{1}{c|}{0.45}              & 0.65           & \multicolumn{1}{c|}{0.63}               & \multicolumn{1}{c|}{0.56}              & \multicolumn{1}{c|}{0.59}              & 0.40           & \multicolumn{1}{c|}{N/A}                & \multicolumn{1}{c|}{N/A}               & \multicolumn{1}{c|}{N/A}               & N/A            \\ \hline
\multicolumn{1}{|c|}{\multirow{3}{*}{\textbf{\begin{tabular}[c]{@{}c@{}}LLM\\ (Agent-based)\end{tabular}}}} & \textbf{\begin{tabular}[c]{@{}c@{}}D-Bot\\ (DeepSeek V3)\end{tabular}}      & \multicolumn{1}{c|}{N/A}                & \multicolumn{1}{c|}{N/A}               & \multicolumn{1}{c|}{N/A}               & N/A            & \multicolumn{1}{c|}{N/A}                & \multicolumn{1}{c|}{N/A}               & \multicolumn{1}{c|}{N/A}               & N/A            & \multicolumn{1}{c|}{0.50}               & \multicolumn{1}{c|}{0.40}              & \multicolumn{1}{c|}{0.45}              & 0.35           & \multicolumn{1}{c|}{N/A}                & \multicolumn{1}{c|}{N/A}               & \multicolumn{1}{c|}{N/A}               & N/A            \\ \cline{2-18} 
\multicolumn{1}{|c|}{}                                                                                      & \textbf{\begin{tabular}[c]{@{}c@{}}D-Bot\\ (DeepSeek-R1 32B)\end{tabular}}  & \multicolumn{1}{c|}{N/A}                & \multicolumn{1}{c|}{N/A}               & \multicolumn{1}{c|}{N/A}               & N/A            & \multicolumn{1}{c|}{N/A}                & \multicolumn{1}{c|}{N/A}               & \multicolumn{1}{c|}{N/A}               & N/A            & \multicolumn{1}{c|}{0.33}               & \multicolumn{1}{c|}{0.33}              & \multicolumn{1}{c|}{0.27}              & 0.50           & \multicolumn{1}{c|}{N/A}                & \multicolumn{1}{c|}{N/A}               & \multicolumn{1}{c|}{N/A}               & N/A            \\ \cline{2-18} 
\multicolumn{1}{|c|}{}                                                                                      & \textbf{\begin{tabular}[c]{@{}c@{}}D-Bot\\ (DeepSeek-R1 671B)\end{tabular}} & \multicolumn{1}{c|}{N/A}                & \multicolumn{1}{c|}{N/A}               & \multicolumn{1}{c|}{N/A}               & N/A            & \multicolumn{1}{c|}{N/A}                & \multicolumn{1}{c|}{N/A}               & \multicolumn{1}{c|}{N/A}               & N/A            & \multicolumn{1}{c|}{0.40}               & \multicolumn{1}{c|}{0.36}              & \multicolumn{1}{c|}{0.34}              & 0.35           & \multicolumn{1}{c|}{N/A}                & \multicolumn{1}{c|}{N/A}               & \multicolumn{1}{c|}{N/A}               & N/A            \\ \hline
\multicolumn{2}{|c|}{\textbf{\begin{tabular}[c]{@{}c@{}}DBAIOps\\ (DeepSeek V3)\end{tabular}}}                                                                                            & \multicolumn{1}{c|}{0.50}               & \multicolumn{1}{c|}{0.67}              & \multicolumn{1}{c|}{0.45}              & 0.66           & \multicolumn{1}{c|}{0.77}               & \multicolumn{1}{c|}{0.87}              & \multicolumn{1}{c|}{\textbf{1.00}}     & 0.88           & \multicolumn{1}{c|}{0.83}               & \multicolumn{1}{c|}{0.91}              & \multicolumn{1}{c|}{0.75}              & 0.83           & \multicolumn{1}{c|}{\textbf{1.00}}      & \multicolumn{1}{c|}{\textbf{1.00}}     & \multicolumn{1}{c|}{0.82}              & \textbf{0.95}  \\ \hline
\multicolumn{2}{|c|}{\textbf{\begin{tabular}[c]{@{}c@{}}DBAIOps\\ (DeepSeek-R1 32B)\end{tabular}}}                                                                                        & \multicolumn{1}{c|}{0.94}               & \multicolumn{1}{c|}{0.88}              & \multicolumn{1}{c|}{0.93}              & 0.87           & \multicolumn{1}{c|}{\textbf{0.94}}      & \multicolumn{1}{c|}{\textbf{0.97}}     & \multicolumn{1}{c|}{\textbf{1.00}}     & 0.95           & \multicolumn{1}{c|}{\textbf{0.87}}      & \multicolumn{1}{c|}{\textbf{0.93}}     & \multicolumn{1}{c|}{\textbf{0.93}}     & 0.85           & \multicolumn{1}{c|}{\textbf{1.00}}      & \multicolumn{1}{c|}{0.95}              & \multicolumn{1}{c|}{\textbf{0.85}}     & 0.90           \\ \hline
\multicolumn{2}{|c|}{\textbf{\begin{tabular}[c]{@{}c@{}}DBAIOps\\ (DeepSeek-R1 671B)\end{tabular}}}                                                                                       & \multicolumn{1}{c|}{\textbf{1.00}}      & \multicolumn{1}{c|}{\textbf{0.95}}     & \multicolumn{1}{c|}{\textbf{1.00}}     & \textbf{0.91}  & \multicolumn{1}{c|}{0.92}               & \multicolumn{1}{c|}{0.96}              & \multicolumn{1}{c|}{\textbf{1.00}}     & \textbf{0.98}  & \multicolumn{1}{c|}{0.83}               & \multicolumn{1}{c|}{0.91}              & \multicolumn{1}{c|}{0.91}              & \textbf{0.88}  & \multicolumn{1}{c|}{\textbf{1.00}}      & \multicolumn{1}{c|}{\textbf{1.00}}     & \multicolumn{1}{c|}{0.82}              & \textbf{0.95}  \\ \hline
\end{tabular}
}
\vspace{1em}
\end{table*}

\noindent \textbf{Evaluation Methods.} We evaluate methods in Table~\ref{tab:relatedwork} that can generate complete diagnosis reports with detailed analysis steps.


\noindent \textsf{(1) Rule-based Tool $+$ DBA:} Utilize pre-defined tools to generate specialized reports, requiring further analysis of an expert DBA to overcome the limitations that traditional methods in Table~\ref{tab:relatedwork} can not generate comprehensive diagnosis reports;

\noindent \textsf{(2) \llm-Only:} We evaluate typical LLMs (DeepSeek-R1-32B and DeepSeek-R1-671B) by directly providing them with the necessary diagnosis information (e.g., the monitoring metrics) as the input.

\noindent \textsf{(3) ChatDBA~\cite{chatdba}:} RAG-based approach that incorporates a tree-based structure to support diagnosis over MySQL and PostgreSQL;

\noindent {\textsf{(4) D-Bot~\cite{dbot2024}:} State-of-the-art \llm-based method that utilizes multi-agent framework (equipped with tree-of-thought algorithm) for diagnosis over PostgreSQL~\cite{dbot2024};

\noindent {\textsf{(5) \oursys:} We provided the metric data and textual knowledge description from the anomaly model and \om knowledge graph, with different \llms as the underlying backbones (i.e., DeepSeek V3~\cite{DeepSeekV3}, DeepSeek-R1-32B, and DeepSeek-R1-671B~\cite{DeepSeekR1});}

Note that we exclude the applications of closed-source LLMs (e.g., GPT-4o~\cite{gpt4o}) from our experiments due to data privacy constraints, as the target database systems reside on private servers. Additionally, conventional ML-based methods are omitted in our evaluation since they only output fixed root cause labels for predefined anomalies without interpretability.
We also exclude \llm-based methods such as Andromeda~\cite{Andromeda}, which focuses on knob recommendation, a different problem from anomaly diagnosis, and GaussMaster~\cite{gaussmaster}, which is limited to the diagnosis of GaussDB, and Panda~\cite{panda}, which lacks open-source code, making comprehensive reproduction of the results infeasible.







\noindent \textbf{Evaluation Metrics.} We adopt four metrics for practical diagnosis evaluation. First, we utilize two basic metrics (i.e., \emph{Precision} and \emph{F1 Score}) to quantify the effectiveness of different methods in root cause identification. Second, we use the metric \emph{Accuracy} proposed in~\cite{dbot2024} to quantify the effectiveness of root cause analysis, considering also the wrong root cause presented. 
The equation of the metric is presented below. 
	$$\text{Acc} = 
	\begin{cases}
	\frac{A_c - \sigma \cdot A_w}{A_a}, & \text{if } A_a > 0 \land A_c \geq \sigma \cdot A_w \\
	0. & \text{otherwise}
	\end{cases}$$
where $A_c$ denotes the number of correct causes, $A_a$ denotes the total number of causes, $A_w$ denotes the number of wrongly detected causes, and $\sigma$ is a hyper-parameter with 0.1 as the default value. 
    
\indent Finally, we adopt Human Evaluation Accuracy (\texttt{HEval}) to measure the overall diagnosis quality of different methods, strictly adhering to three human-assessed criteria$\footnote{\url{https://github.com/weAIDB/DBAIOps/blob/master/HEval_criteria.md}}$.
\textsf{(1) Root Cause Recall (30\%):} Whether identify all the relevant root causes;
\textsf{(2) Theoretical Consistency (30\%):} Whether the diagnosis reasoning aligns with theoretical knowledge logically (e.g., adhering to Oracle database principles and operating system mechanisms);
\textsf{(3) Evidence Authenticity (40\%):} Whether the evidence data (e.g., redo log write times, storage latency metrics) supporting the diagnosis reasoning is valid and free from hallucinations.


\noindent \textbf{Other Settings.}
The experimental setup includes the following key components: (1) LLM SERVER, utilizing the Ollama framework and equipped with an RTX 3090 GPU, running a 32B distilled model; 
(2) An operational knowledge graph, constructed based on the KYD Zhiyan platform; and (3) Data collection, performed using the \oursys~community edition tool. 
These components collectively provide the necessary technical support for the experiment, ensuring efficient operation and accurate data analysis.
(4) Database, We implemented \oursys and conducted experiments on four databases: Oracle\cite{Oracle}, MySQL\cite{MySQL}, PostgreSQL\cite{PostgreSQL}, and Dameng Database (DM8)\cite{Dameng}.

\subsection{Overall Performance}
\label{subsub:overall}

We assess and compare the effectiveness of different methods over anomalies across diverse database systems, and Table~\ref{tab:overall} presents the overall diagnosis performance.
Based on the experimental results, we have the following observations.

First, \emph{\oursys~achieves comparable diagnosis performance under different \llms, with {highest} performance over the four database systems.} Specifically, \oursys(DeepSeek-R1 32B) and \oursys(DeepSeek-R1 671B) obtain the aggregated diagnosis performance of {0.92} and {0.94}, which is {61.40\%} and {34.29\%} higher than the diagnosis performance of {0.57} by DeepSeek-R1 32B and {0.70} by DeepSeek-R1 671B, respectively.
The underlying reason can be attributed to the fact that even with the necessary diagnosis information (e.g., the relevant metrics), \llm~can only conclude the root causes based on their general knowledge rather than specific \om experience in the graph model of \oursys.

\begin{figure*}[!t]
    \centering
    \begin{subfigure}{.49\textwidth}
        \includegraphics*[width=\textwidth]{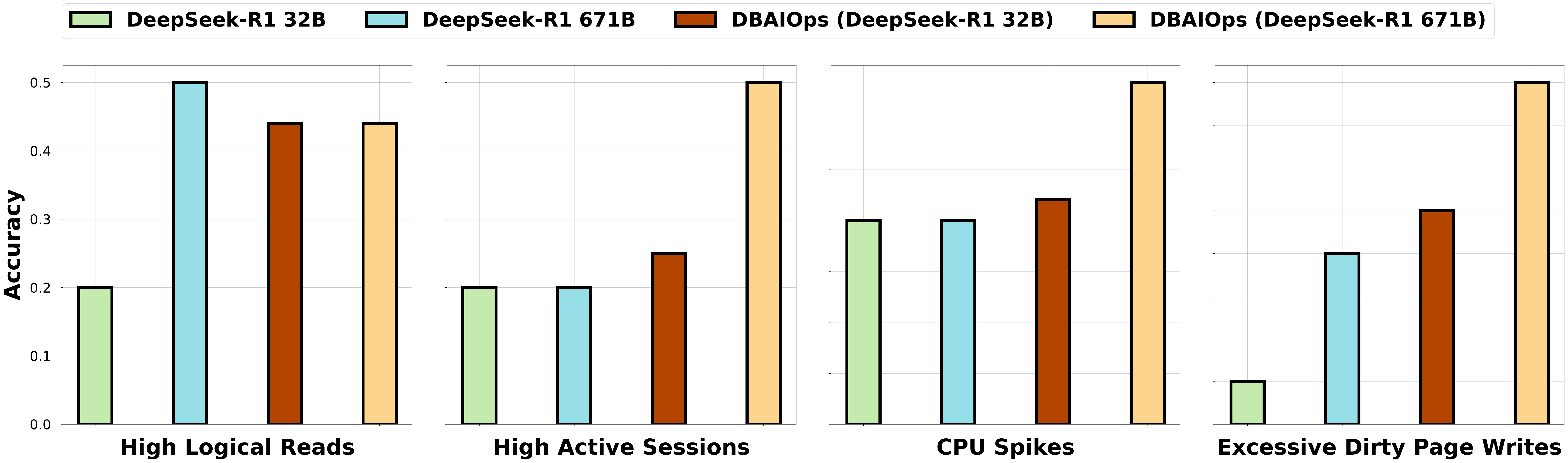}
        \caption{Precision}
        \label{fig:overall_acc}
    \end{subfigure}
    \begin{subfigure}{.49\textwidth}
        \includegraphics*[width=\textwidth]{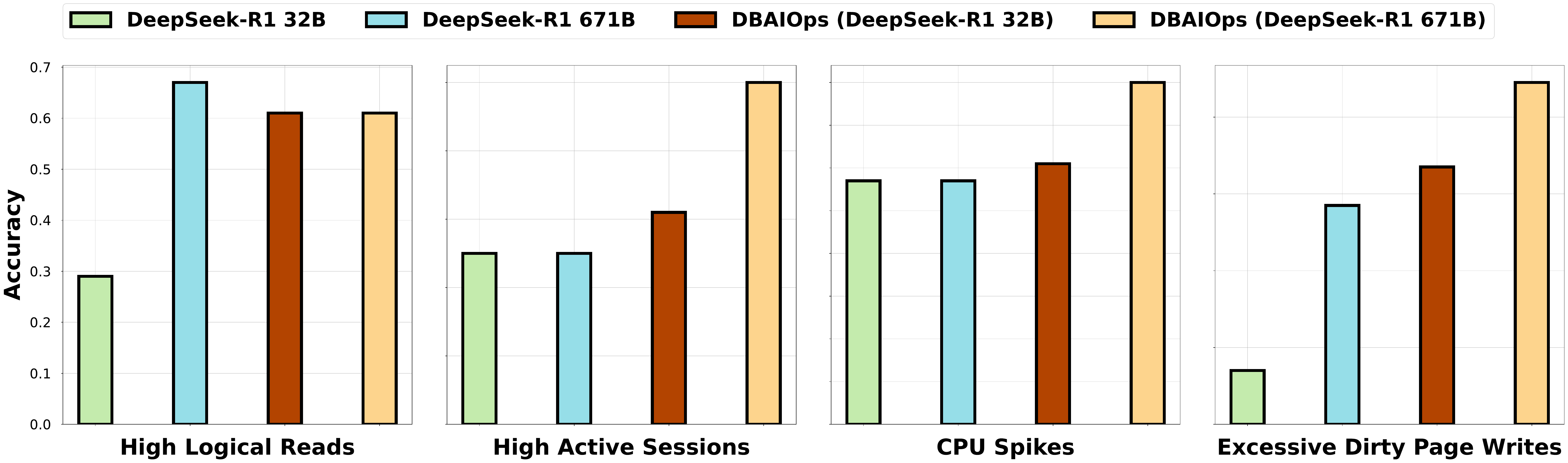}
        \caption{F1-Score}
        \label{fig:overall_HEval}
    \end{subfigure}
    \begin{subfigure}{.49\textwidth}
        \includegraphics*[width=\textwidth]{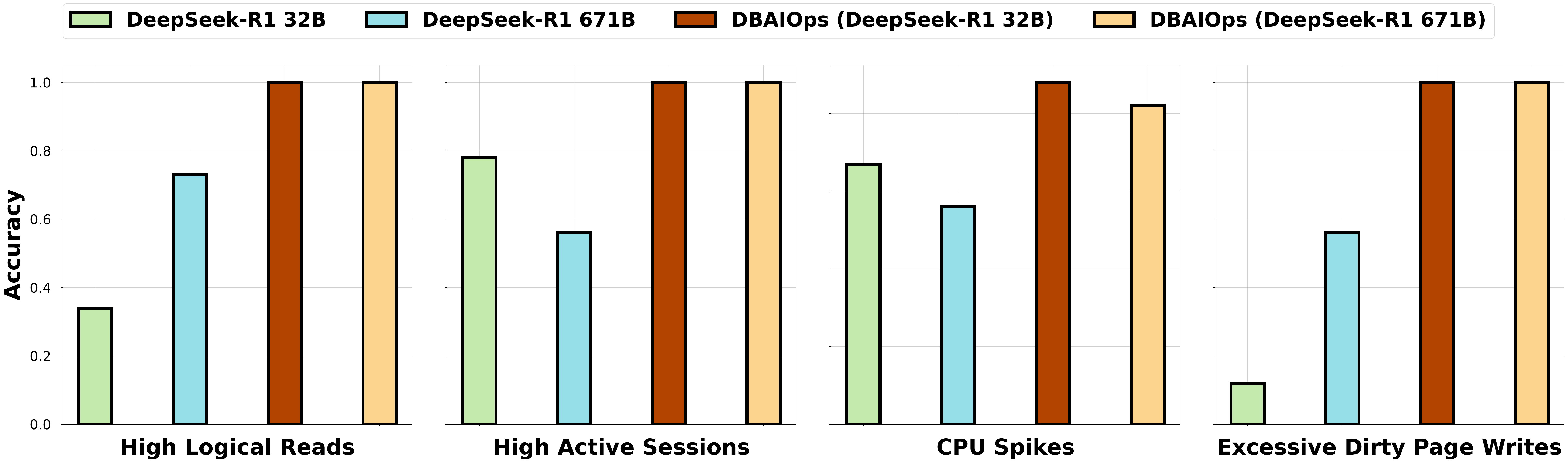}
        \caption{Accuracy}
        \label{fig:overall_acc}
    \end{subfigure}
    \begin{subfigure}{.49\textwidth}
        \includegraphics*[width=\textwidth]{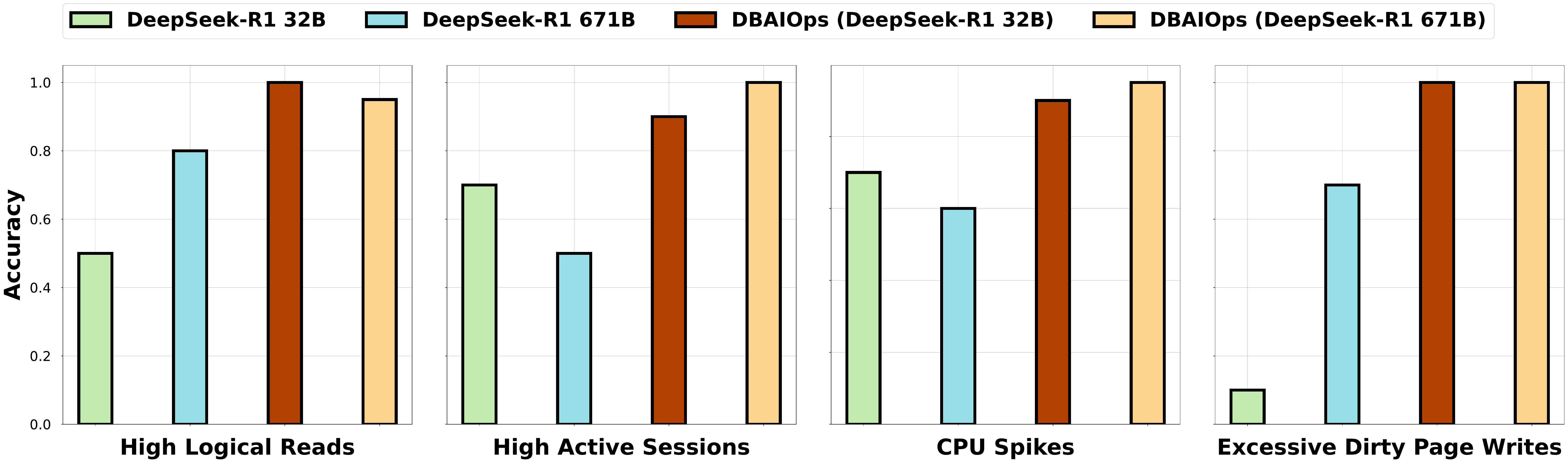}
        \caption{HEval}
        \label{fig:overall_HEval}
    \end{subfigure}
    \caption{Distribution of Result Accuracy and HEval across Different Scenarios.}
    \label{fig:overall}
\end{figure*}

Second, \emph{\oursys can {outperform state-of-the-art \llm-based methods} or even \textsf{Rule-based Tool $+$ DBA} method, showcasing well-behaved generalization ability across scenarios.} Specifically, {\oursys outperforms D-Bot and ChatDBA by over 37\% and 45\% in \emph{HEval}}.
The average diagnosis performance of \oursys~with different \llm arrives at {0.89} across database systems, which is comparable with the performance of {0.91} by \textsf{Rule-based Tool $+$ DBA}.
{Moreover, \textsf{\oursys} obtains the average accuracy of 0.94, better than 0.79 of \textsf{Rule-based Tool $+$ DBA} over MySQL anomalies. The underlying reason is that DBAs face challenges in processing large volumes of monitoring data within a limited time.
They often rely on a small subset of signals, which can lead to incomplete or conflicting conclusions.
For example, in \emph{IO\_Latency\_MySQL\_Critical} anomaly model, the tool generates 14 diagnosis items, making it difficult for DBAs to analyze all relevant data and accurately identify all root causes.}
In contrast, \oursys~improves in two aspects: (1) \oursys~provides \llm with more comprehensive \om experience essential for accurate diagnosis (e.g., metric statistics and relevant knowledge points), some of which might be left out in the pertaining corpus of \llm;
(2) \oursys~carefully prompts them to reason over the provided information about the anomaly like DBA (e.g., analyze the provided metrics via relevant \om experience) and exploits the generation capability of \llms to produce customized diagnosis reports.
Thus, \oursys can generate more comprehensive and user-friendly (i.e., easier to understand) diagnosis reports (More details in Section~\ref{subsec:report}).

Finally, \emph{\oursys~with the medium-sized reasoning model, i.e., \textsf{\oursys(DeepSeek-R1 32B)} can achieve comparable diagnosis accuracy to a large-scale reasoning model, i.e., \textsf{\oursys(DeepSeek-R1 671B)}.}
Specifically, the average diagnosis performance of \textsf{\oursys(DeepSeek-R1 32B)} arrives at {0.92} across database systems, comparable to the one of {0.94} by \textsf{\oursys(DeepSeek-R1 671B)}.
The reason is that \oursys~provides useful information, including in the processed data from the combined usage of anomaly model and \om knowledge graph, alleviating the difficulty in identifying the correct root causes.
Therefore, a medium-sized reasoning model can behave well based on the information in the processed data.

\begin{table}[!t]
\caption{Performance of \oursys~Variants.}
\label{tab:ablation}
\resizebox{\linewidth}{!}{
\begin{tabular}{|c|c|c|c|c|}
\hline
\textbf{Method}                                                                   & \textbf{Precision} & \textbf{F1-Score} & \textbf{Accuracy} & \textbf{HEval} \\ \hline
\textbf{w/o ADF}                                                     & 0.75               & 0.86              & 0.73              & 0.50            \\ \hline
\textbf{w/o Graph Evolution}                                                                  & 0.67               & 0.80               & 0.63              & 0.60            \\ \hline
\textbf{\begin{tabular}[c]{@{}c@{}}w/o ADF + \\w/o Graph Evolution\end{tabular}} & 0.60               & 0.75              & 0.56              & 0.50            \\ \hline
\end{tabular}
}
\vspace{-.5cm}
\end{table}

We further investigate the detailed diagnosis performance of different methods across database anomalies.
Figure~\ref{fig:overall} presents the results of diagnosis performance over four scenarios.
We notice that \emph{\oursys~showcases stable diagnosis performance improvement of \llms that present different effectiveness across anomalies.} 
Specifically, \oursys~helps to achieve the promising diagnosis performance of {0.87} on average across different anomalies.
In contrast, \textsf{DeepSeek-R1 32B} performs well with the average diagnosis performance of {0.59} over {CPU Spikes} and performs poorly with the average diagnosis performance of {0.12} over {Excessive Dirty Page Writes}.
It reflects that solely relying on the internal \om experience for accurate diagnosis over different anomalies might be problematic, making the paradigm of augmenting experience with \om knowledge graph in \oursys a necessity.

\begin{table}[!t]
\caption{Similarity Analysis of \llm Outputs.}
\label{tab:output_similarity}
\resizebox{\linewidth}{!}{
\begin{tabular}{|c|cccc|cccc|}
\hline
\multirow{2}{*}{\textbf{Method}}                                              & \multicolumn{4}{c|}{\textbf{PostgreSQL}}                                                                                                    & \multicolumn{4}{c|}{\textbf{Oracle}}                                                                                                        \\ \cline{2-9} 
                                                                              & \multicolumn{1}{c|}{\textbf{difflib}} & \multicolumn{1}{c|}{\textbf{Levenshtein}} & \multicolumn{1}{c|}{\textbf{Jaccard}} & \textbf{TF-IDF} & \multicolumn{1}{c|}{\textbf{difflib}} & \multicolumn{1}{c|}{\textbf{Levenshtein}} & \multicolumn{1}{c|}{\textbf{Jaccard}} & \textbf{TF-IDF} \\ \hline
\textbf{\begin{tabular}[c]{@{}c@{}}\oursys\\ (DeepSeek-R1 32B)\end{tabular}}  & \multicolumn{1}{c|}{0.04}             & \multicolumn{1}{c|}{0.07}                 & \multicolumn{1}{c|}{0.20}             & 0.30            & \multicolumn{1}{c|}{0.04}             & \multicolumn{1}{c|}{0.05}                 & \multicolumn{1}{c|}{0.15}             & 0.39            \\ \hline
\textbf{\begin{tabular}[c]{@{}c@{}}\oursys\\ (DeepSeek-R1 671B)\end{tabular}} & \multicolumn{1}{c|}{0.04}             & \multicolumn{1}{c|}{0.10}                 & \multicolumn{1}{c|}{0.24}             & 0.37            & \multicolumn{1}{c|}{0.05}             & \multicolumn{1}{c|}{0.09}                 & \multicolumn{1}{c|}{0.29}             & 0.50            \\ \hline
\end{tabular}
}
\vspace{-.35cm}
\end{table}



\begin{table*}[!t]
\caption{Case Study of Diagnosis Reports Generated by \oursys and Baseline Methods (Complete reports are in our \href{https://github.com/weAIDB/DBAIOps/blob/master/Appendix_Diagnosis_Report.pdf}{\gray{[\underline{artifact}]}}).}
\vspace{-.2cm}
\label{tab:report}
\resizebox{\linewidth}{!}{
\begin{tabular}{|cc|ll|ll|}
\hline
\multicolumn{2}{|c|}{}                                                                                                                                                                 & \multicolumn{2}{c|}{\textbf{\begin{tabular}[c]{@{}c@{}}\textsf{LOG SYNCHRONIZATION DELAY}\\ (Oracle Anomaly)\end{tabular}}}                                                                                                                                                                                                                                                                                                                                                                                                                                                                                     & \multicolumn{2}{c|}{\textbf{\begin{tabular}[c]{@{}c@{}}\textsf{BACKEND PROCESS FLUSHES DIRTY PAGES}\\ (PostgreSQL Anomaly)\end{tabular}}}                                                                                                                                                                                                                                                                                                                                                                                                                                                                                                                                                                                                      \\ \cline{3-6} 
\multicolumn{2}{|c|}{\multirow{-3}{*}{\textbf{\begin{tabular}[c]{@{}c@{}}HEval\\ Criteria\end{tabular}}}}                                                                                                                                                                 & \multicolumn{2}{l|}{\begin{tabular}[c]{@{}l@{}}\textbf{Anomaly Description:}\\ The wait occurs during commits or rollbacks while waiting for redo logs to be written to disk, \\ often causing bottlenecks under heavy transactions or poor I/O.\end{tabular}}                                                                                                                                                                                                                                                                                                                                                  & \multicolumn{2}{l|}{\begin{tabular}[c]{@{}l@{}}\textbf{Anomaly Description:}\\ The abnormal alarm on the metric backend buffer write ratio, which indicates \\ potential misconfigurations in shared buffers or bgwriter/checkpoint settings.\end{tabular}}                                                                                                                                                                                                                                                                                                                                                                                                                                                                                    \\ \cline{3-6} 
\multicolumn{2}{|c|}{}                                                                              & \multicolumn{1}{c|}{\textbf{\begin{tabular}[c]{@{}c@{}}Positive Example\\ (HEval = 1.00, by \oursys)\end{tabular}}}                                                                                                                                                                             & \multicolumn{1}{c|}{\textbf{\begin{tabular}[c]{@{}c@{}}Negative Example\\ (HEval = 0.40, by Baselines)\end{tabular}}}                                                                                                                                                                                           & \multicolumn{1}{c|}{\textbf{\begin{tabular}[c]{@{}c@{}}Positive Example\\ (HEval = 1.00, by \oursys)\end{tabular}}}                                                                                                                                                                                          & \multicolumn{1}{c|}{\textbf{\begin{tabular}[c]{@{}c@{}}Negative Example\\ (HEval = 0.00, by Baselines)\end{tabular}}}                                                                                                                                                                                                                                                                                                                            \\ \hline
\multicolumn{1}{|c|}{\textbf{\begin{tabular}[c]{@{}c@{}}Root Cause\\ Recall\end{tabular}}}                                                                                             & \textbf{\begin{tabular}[c]{@{}c@{}}Report\\ Content\end{tabular}} & \multicolumn{1}{l|}{\cellcolor[HTML]{9AFF99}\begin{tabular}[c]{@{}l@{}} \ding{52} \textbf{Root Cause:}\\ • (1) Insufficient I/O Performance of REDO Log Storage\\ • (2) Intermittent I/O Pressure Spikes during Log Writing\end{tabular}}                                   & \cellcolor[HTML]{FFCCC9}\begin{tabular}[c]{@{}l@{}} \ding{56} \textbf{Root Cause:}\\ • (1) Log file parallel write anomaly\\ • (2) Redo generation rate anomaly\\ • (3) Checkpoint delay anomaly\\ • (4) Insufficient memory\\ • (5) Control file write anomaly\end{tabular} & \multicolumn{1}{l|}{\cellcolor[HTML]{9AFF99}\begin{tabular}[c]{@{}l@{}} \ding{52} \textbf{Root Cause:}\\ • (1) bgwriter\_lru\_maxpages too low\\ • (2) I/O latency causing bgwriter failure\\ • (3) bgwriter\_lru\_multiplier too low\end{tabular}}                       & \cellcolor[HTML]{FFCCC9}\begin{tabular}[c]{@{}l@{}} \ding{56} \textbf{Root Cause:}\\ • (1) High I/O latency\\ • (2) Misconfigured checkpoints\\ • (3) Insufficient bgwriter\\ • (4) High concurrent writes\\ • (5) Unoptimized SQL\end{tabular}                                                                                                                                                                 \\ \cline{2-6} 
\multicolumn{1}{|c|}{\multirow{-2}{*}{}}       & \textbf{Comment}                                                  & \multicolumn{1}{l|}{\begin{tabular}[c]{@{}l@{}}• Identify both root causes centered \\ around I/O storage bottleneck\end{tabular}}                                                                                                                                                                             & \begin{tabular}[c]{@{}l@{}}• Only list symptoms and manifestations\\ (i.e., root cause (1),(3), and (5))\\ • Miss core root cause around I/O bottleneck\\ (i.e., (2) and (4) are not direct root causes)\end{tabular}                                                                                           & \multicolumn{1}{l|}{\begin{tabular}[c]{@{}l@{}}• Correctly center around root causes of\\ BGWRITER parameters and I/O bottlenecks\end{tabular}}                                                                                                                                                              & \begin{tabular}[c]{@{}l@{}}• Root cause (4) and (5) describe transaction backlog,\\ irrelevant to REDO logs\\ • Root cause (1) - (3) partially valid but diluted\end{tabular}                                                                                                                                                                                                                                                                    \\ \hline
\multicolumn{1}{|c|}{\textbf{\begin{tabular}[c]{@{}c@{}}Theoretical\\ Consistency\end{tabular}}}                                                                                             & \textbf{\begin{tabular}[c]{@{}c@{}}Report\\ Content\end{tabular}} & \multicolumn{1}{l|}{\cellcolor[HTML]{9AFF99}\begin{tabular}[c]{@{}l@{}} \ding{52} \textbf{Reasoning:} \\ • log\_file\_sync / log\_file\_parallel\_write = 2\\    → Storage I/O primary factor\\ • Spike at 06:00 with normal OS latency\\    → transient load\end{tabular}} & \cellcolor[HTML]{9AFF99}\begin{tabular}[c]{@{}l@{}} \ding{52} \textbf{Reasoning:}\\ • Redo surge \\    → log buffer overflow \\    → wait for LGWR\\ • Memory pressure\\    → log buffer insufficient\end{tabular}                                                           & \multicolumn{1}{l|}{\cellcolor[HTML]{9AFF99}\begin{tabular}[c]{@{}l@{}} \ding{52} \textbf{Reasoning:}\\ • bgwriter\_stop\_scan\_count \textgreater{}0 \\    → reach bgwriter\_lru\_maxpages limit\\       → bgwriter stops\\    → backend takes over writes\end{tabular}} & \cellcolor[HTML]{FFCCC9}\begin{tabular}[c]{@{}l@{}} \ding{56} \textbf{Reasoning:}\\ • I/O latency had a maximum value of 5736.96ms\\   and an average value of 827.0ms\\   (normal value should be \textless{}10ms)\\ • active sessions had an average of 48\\    → high concurrency\\ • checkpoint delay had a maximum value \\   of 525,058,688.0ms\\    → checkpoint process severely blocked\end{tabular} \\ \cline{2-6} 
\multicolumn{1}{|c|}{\multirow{-2}{*}{}} & \textbf{Comment}                                                  & \multicolumn{1}{l|}{\begin{tabular}[c]{@{}l@{}}• Rigorous reasoning chain linking metrics \\ to Oracle and OS knowledge or mechanisms\end{tabular}}                                                                                                                                                            & \begin{tabular}[c]{@{}l@{}}• Despite root cause errors,\\ reasoning adheres to Oracle principles\end{tabular}                                                                                                                                                                                                   & \multicolumn{1}{l|}{• Logically clear, aligned with OS knowledge}                                                                                                                                                                                                                                            & \begin{tabular}[c]{@{}l@{}}• Merely describe abnormal metric alarms\\ rather than establishing causal relations\end{tabular}                                                                                                                                                                                                                                                                                                                     \\ \hline
\multicolumn{1}{|c|}{\textbf{\begin{tabular}[c]{@{}c@{}}Evidence\\ Authenticity\end{tabular}}}                                                                                             & \textbf{\begin{tabular}[c]{@{}c@{}}Report\\ Content\end{tabular}} & \multicolumn{1}{l|}{\cellcolor[HTML]{9AFF99}\begin{tabular}[c]{@{}l@{}} \ding{52} \textbf{Evidence:}\\ • Metric 2184301 (log file sync): \\   max=15.2ms, avg=6.0ms\\ • Metric 2184305 (log file parallel write):\\   max=7.09ms, avg=3.0ms\end{tabular}}                   & \cellcolor[HTML]{FFCCC9}\begin{tabular}[c]{@{}l@{}} \ding{56} \textbf{Evidence:}\\ • Metric 2180503 (checkpoint delay):\\   max=61,660.0ms, avg=61,060.0ms\\ • Metric 2184306 (control file write):\\   max=3.78ms, avg=1.0ms\end{tabular}                                   & \multicolumn{1}{l|}{\cellcolor[HTML]{9AFF99}\begin{tabular}[c]{@{}l@{}} \ding{52} \textbf{Evidence:}\\ • Metric 2300140 (Bgwriter stop scan):\\   max=2.76, avg=1.0\\ • Metric 3000006 (I/O latency):\\   max=5736.96ms, avg=827.0ms\end{tabular}}                        & \cellcolor[HTML]{9AFF99}\begin{tabular}[c]{@{}l@{}} \ding{52} \textbf{Evidence:}\\ • Metric 3000006 (I/O latency):\\   max=5736.96ms, avg=827.0ms\\ • Metric 2300145 (checkpoint delay):\\   max=525,058,688.0ms\end{tabular}                                                                                                                                                                                 \\ \cline{2-6} 
\multicolumn{1}{|c|}{\multirow{-2}{*}{}}   & \textbf{Comment}                                                  & \multicolumn{1}{l|}{\begin{tabular}[c]{@{}l@{}}• All metrics exist in provided data\\ (i.e., no hallucination)\end{tabular}}                                                                                                                                                                                   & \begin{tabular}[c]{@{}l@{}}• Cite data that does not exist in provided metrics\\ (i.e., hallucinated metrics)\end{tabular}                                                                                                                                                                                      & \multicolumn{1}{l|}{\begin{tabular}[c]{@{}l@{}}• Metrics match the provided data\\ (i.e., no hallucination)\end{tabular}}                                                                                                                                                                                    & \begin{tabular}[c]{@{}l@{}}• Despite wrong conclusions,\\ metrics exist in provided data\\ (i.e., no hallucination)\end{tabular}                                                                                                                                                                                                                                                                                                                 \\ \hline
\end{tabular}
}
\end{table*}

\subsection{Ablation Study}
\label{subsub:ablation}

We experiment with the following \oursys~variants to investigate the importance of the \om experience. \textsf{(1) w/o ADF:} We adopt a set of fixed thresholds for each metric without the dynamic updating by the ADF algorithm introduced in Section~\ref{subsec:evolution};
\textsf{(2) w/o Graph Evolution:} We remove the two-step graph evolution strategy introduced in Section~\ref{subsec:evolution} with less \om experience accumulated.
\textsf{(3) w/o ADF + Graph Evolution:} We simultaneously remove the above two mechanism altogether. The experiments are on Oracle and the underlying \llm is DeepSeek-R1 32B.


As shown in Table~\ref{tab:ablation}, we observe that \emph{the removal of both the two mechanisms leads to the degration of diagnosis performance of \oursys}.
Specifically, the three variants of \oursys obtains the diagnosis performance of {0.66} on average, which is {34\%} lower than the original version of \oursys.
Moreover, the diagnosis performance of \textsf{w/o Graph Evolution} method is {0.68} on average, which is worse than the \textsf{w/o ADF} of {0.71} on average.
All of these results indicate the importance of \om experience integration for better diagnosis performance.
{The \om graph model enriches \llm with essential, context-specific information (e.g., background on LOG SYNC FILE \emph{Trigger} vertex and dozens of abnormal metric patterns).}
Thus, methods that restrict the graph without sufficient experience lead to diagnosis mistakes.
The \textsf{w/o ADF} method fails to adaptively update the metric threshold to detect the abnormal ones to accumulate more \om experience since the collection is terminated once all the metrics are denoted to be normal.
The \textsf{w/o Graph Evolution} method strictly restricts the collection of \om experience in the neighborhood without the consideration of relevant anomalies, typically leading to a higher volume of \om experience loss.
Thus, without the specific \om experience, theses methods is difficult to accurately pinpoint the underlying root causes and the recovery solutions for better diagnosis performance.


{We also assess the similarity between \llm outputs and input graph content using character-level (i.e., Levenshtein distance~\cite{LevenshteinDistance}) and word-level (i.e., TF-IDF~\cite{TFIDF}) metrics.
As shown in Table~\ref{tab:output_similarity}, for PostgreSQL anomalies, we obtain similarity scores of 0.10 and 0.37, showing that the outputs are not merely copied from the graph.}

\subsection{Real-World Case Analysis}
\label{subsec:report}


We assess the diagnosis reports generated by \oursys and baseline methods for two representative anomalies (i.e., \textsf{LOG SYNCHRONIZATION DELAY} in Oracle, and \textsf{BACKEND PROCESS FLUSHES DIRTY PAGES} in PostgreSQL).
As summarized in Table~\ref{tab:report}, we evaluate two diagnosis reports for each anomaly using the three criteria (i.e., Root Cause Recall, Theoretical Consistency, and Evidence Authenticity) of Human Evaluation Accuracy (\texttt{HEval}) in Section~\ref{subsec:anomaly}.

\noindent $\blacktriangleright$ \bfit{Root Cause Recall.}
This criterion examines whether the reports correctly identify the direct root causes of anomalies.
As shown in Table~\ref{tab:report}, \oursys~accurately captures all root causes related to I/O bottlenecks and parameter misconfigurations (e.g., correctly identifying that \textsf{bgwriter\_lru\_maxpages} is set too low).
However, the baseline method mainly lists secondary symptoms rather than direct causes.
For instance, in Oracle anomaly, it only reports abnormal patterns of log file parallel write, checkpoint delay, and control file write, missing the actual root cause (i.e., insufficient I/O performance of REDO log storage).
This improvement stems from the experience graph and anomaly model in \oursys, which effectively characterize and supply \llm with essential knowledge for comprehensive diagnosis (Section~\ref{sec:graph} and Section~\ref{sec:anomaly}).


\noindent $\blacktriangleright$ \bfit{Theoretical Consistency.}
This criterion evaluates whether the reasoning follows established database principles.
As displayed in Table~\ref{tab:report}, \oursys consistently grounds its reasoning in theoretical knowledge, forming causal chains that link abnormal metrics to underlying mechanisms.
For example, it explicitly relates abnormal \textsf{bgwriter stop scan} counts to \textsf{bgwriter\_lru\_maxpages} and \textsf{bgwriter\_lru\_multiplier} settings based on bgwriter and checkpoint principles.
In contrast, the baseline method merely enumerates abnormal alarms (e.g., elevated I/O latency and active sessions) without constructing causal relations to arrive at the root causes (i.e., misconfiguration of \textsf{bgwriter\_lru\_maxpages} parameter).
This strength arises from the knowledge-path-based diagnosis strategy in \oursys, which augments \llms with well-organized diagnosis context, and steers them with structured prompts to reason over the path carefully (Section~\ref{subsec:reasoning}).

\noindent $\blacktriangleright$ \bfit{Evidence Authenticity.}
This criterion assesses whether all cited evidence originates from the provided data.
As shown in Table~\ref{tab:report}, \oursys~exclusively relies on the provided evidence data (e.g., \textsf{log file parallel write} in Oracle and \textsf{I/O latency} in PostgreSQL), while the baseline occasionally cites non-existent or incorrect values, such as a non-existent checkpoint delay metric and an inaccurate maximum of 3.78ms for control file write due to hallucination issues that compromise reliability. Thus, \oursys can better constrain the \llm to reason strictly based on the essential evidence identified through the graph and its automatic evolution mechanism (Section~\ref{subsec:evolution}).



%% file: 9_conclusion.tex
\section{Conclusion}

\begin{sloppypar}
In this paper, we presented the first hybrid database \om system \oursys, which combines the benefits of knowledge graphs and reasoning \llms to support real-world \om for 25 databases, covering domains like finance and healthcare.
We constructed a heterogeneous graph model that enables the reuse of structured \om experience across different database systems.
We designed a collection of anomaly models from a fine-grained metric hierarchy that captures explicit and implicit metric correlations.  
We propose a two-stage graph evolution mechanism to adaptively explore diagnosis paths and accumulate experiences for newly observed anomalies.
We introduced a long-term reasoning mechanism that guides diagnosis through the context of adaptive graph traversal and LLM-based inference.
Extensive experiments validated the effectiveness of \oursys, demonstrating superior root cause accuracy and report quality compared to traditional and \llm-based methods.
\end{sloppypar}




%% file: main.bbl

\begin{thebibliography}{49}


\ifx \showCODEN    \undefined \def \showCODEN     #1{\unskip}     \fi
\ifx \showDOI      \undefined \def \showDOI       #1{#1}\fi
\ifx \showISBNx    \undefined \def \showISBNx     #1{\unskip}     \fi
\ifx \showISBNxiii \undefined \def \showISBNxiii  #1{\unskip}     \fi
\ifx \showISSN     \undefined \def \showISSN      #1{\unskip}     \fi
\ifx \showLCCN     \undefined \def \showLCCN      #1{\unskip}     \fi
\ifx \shownote     \undefined \def \shownote      #1{#1}          \fi
\ifx \showarticletitle \undefined \def \showarticletitle #1{#1}   \fi
\ifx \showURL      \undefined \def \showURL       {\relax}        \fi
\providecommand\bibfield[2]{#2}
\providecommand\bibinfo[2]{#2}
\providecommand\natexlab[1]{#1}
\providecommand\showeprint[2][]{arXiv:#2}

\bibitem[\protect\citeauthoryear{??}{DBA}{[n.d.]}]%
        {DBAExchange}
 \bibinfo{year}{[n.d.]}\natexlab{}.
\newblock \showarticletitle{https://dba.stackexchange.com/}.
\newblock
\newblock
\shownote{Last accessed on 2025-07.}


\bibitem[\protect\citeauthoryear{??}{MyS}{[n.d.]}]%
        {MySQLForum}
 \bibinfo{year}{[n.d.]}\natexlab{}.
\newblock \showarticletitle{https://forums.mysql.com/}.
\newblock
\newblock
\shownote{Last accessed on 2025-07.}


\bibitem[\protect\citeauthoryear{??}{SQL}{[n.d.]}]%
        {SQLServerQA}
 \bibinfo{year}{[n.d.]}\natexlab{}.
\newblock \showarticletitle{https://learn.microsoft.com/en-us/answers/tags/780/sql-server}.
\newblock
\newblock
\shownote{Last accessed on 2025-07.}


\bibitem[\protect\citeauthoryear{??}{gpt}{[n.d.]}]%
        {gpt4o}
 \bibinfo{year}{[n.d.]}\natexlab{}.
\newblock \showarticletitle{https://openai.com/index/hello-gpt-4o/}.
\newblock
\newblock
\shownote{Last accessed on 2024-10.}


\bibitem[\protect\citeauthoryear{??}{pro}{[n.d.]}]%
        {prometheus}
 \bibinfo{year}{[n.d.]}\natexlab{}.
\newblock \showarticletitle{https://prometheus.io/}.
\newblock
\newblock
\shownote{Last accessed on 2025-07.}


\bibitem[\protect\citeauthoryear{??}{MOS}{[n.d.]}]%
        {MOS}
 \bibinfo{year}{[n.d.]}\natexlab{}.
\newblock \showarticletitle{https://support.oracle.com}.
\newblock
\newblock
\shownote{Last accessed on 2025-04.}


\bibitem[\protect\citeauthoryear{??}{Neo}{[n.d.]}]%
        {Neo4j}
 \bibinfo{year}{[n.d.]}\natexlab{}.
\newblock \showarticletitle{https://www.neo4j.com/}.
\newblock
\newblock
\shownote{Last accessed on 2025-07.}


\bibitem[\protect\citeauthoryear{??}{cha}{[n.d.]}]%
        {chatdba}
 \bibinfo{year}{[n.d.]}\natexlab{}.
\newblock \showarticletitle{http://web.chatdba.com/}.
\newblock
\newblock
\shownote{Last accessed on 2025-04.}


\bibitem[\protect\citeauthoryear{??}{FAA}{tage}]%
        {FAA1}
 \bibinfo{year}{2023 FAA system outage}\natexlab{}.
\newblock \bibinfo{booktitle}{\emph{(Wikipedia)}}.
\newblock
\urldef\tempurl%
\url{https://en.wikipedia.org/wiki/2023_FAA_system_outage}
\showURL{%
\tempurl}
\newblock
\shownote{Last accessed on 2025-07.}


\bibitem[\protect\citeauthoryear{??}{Ali}{base}]%
        {AlibabaDatabase}
 \bibinfo{year}{Alibaba Database}\natexlab{}.
\newblock \bibinfo{booktitle}{\emph{(Database)}}.
\newblock
\urldef\tempurl%
\url{https://www.alibabacloud.com/en/product/databases}
\showURL{%
\tempurl}


\bibitem[\protect\citeauthoryear{??}{Ama}{base}]%
        {AmazonDatabase}
 \bibinfo{year}{Amazon Database}\natexlab{}.
\newblock \bibinfo{booktitle}{\emph{(Database)}}.
\newblock
\urldef\tempurl%
\url{https://aws.amazon.com/cn/free/database}
\showURL{%
\tempurl}


\bibitem[\protect\citeauthoryear{??}{Dam}{base}]%
        {Dameng}
 \bibinfo{year}{Dameng Database}\natexlab{}.
\newblock \bibinfo{booktitle}{\emph{(DBMS)}}.
\newblock
\urldef\tempurl%
\url{https://en.dameng.com/}
\showURL{%
\tempurl}


\bibitem[\protect\citeauthoryear{??}{dat}{Vail}]%
        {dataVail}
 \bibinfo{year}{dataVail}\natexlab{}.
\newblock \bibinfo{booktitle}{\emph{(DBMS)}}.
\newblock
\urldef\tempurl%
\url{https://www.datavail.com/solutions/database-administration/}
\showURL{%
\tempurl}


\bibitem[\protect\citeauthoryear{??}{FAA}{ghts}]%
        {FAA2}
 \bibinfo{year}{FAA says ‘damaged database file’ prompted halt on domestic US flights}\natexlab{}.
\newblock \bibinfo{booktitle}{\emph{(News)}}.
\newblock
\urldef\tempurl%
\url{https://www.ft.com/content/e65ee681-f242-45f1-b1ab-b5f1b42d8a12}
\showURL{%
\tempurl}
\newblock
\shownote{Last accessed on 2025-07.}


\bibitem[\protect\citeauthoryear{??}{MyS}{ySQL}]%
        {MySQL}
 \bibinfo{year}{MySQL}\natexlab{}.
\newblock \bibinfo{booktitle}{\emph{(DBMS)}}.
\newblock
\urldef\tempurl%
\url{https://www.mysql.com/}
\showURL{%
\tempurl}


\bibitem[\protect\citeauthoryear{??}{Ora}{acle}]%
        {Oracle}
 \bibinfo{year}{Oracle}\natexlab{}.
\newblock \bibinfo{booktitle}{\emph{(DBMS)}}.
\newblock
\urldef\tempurl%
\url{https://www.oracle.com/database/}
\showURL{%
\tempurl}


\bibitem[\protect\citeauthoryear{??}{Per}{cona}]%
        {Percona}
 \bibinfo{year}{Percona}\natexlab{}.
\newblock \bibinfo{booktitle}{\emph{(DBMS)}}.
\newblock
\urldef\tempurl%
\url{https://try.percona.com/managed-services/}
\showURL{%
\tempurl}


\bibitem[\protect\citeauthoryear{??}{Pos}{eSQL}]%
        {PostgreSQL}
 \bibinfo{year}{PostgreSQL}\natexlab{}.
\newblock \bibinfo{booktitle}{\emph{(DBMS)}}.
\newblock
\urldef\tempurl%
\url{https://www.postgresql.org}
\showURL{%
\tempurl}


\bibitem[\protect\citeauthoryear{??}{Rac}{pace}]%
        {Rackspace}
 \bibinfo{year}{Rackspace}\natexlab{}.
\newblock \bibinfo{booktitle}{\emph{(DBMS)}}.
\newblock
\urldef\tempurl%
\url{https://docs.rackspace.com/docs/database-administration-solutions/}
\showURL{%
\tempurl}


\bibitem[\protect\citeauthoryear{Benoit}{Benoit}{2005}]%
        {ADPP}
\bibfield{author}{\bibinfo{person}{Darcy~G. Benoit}.} \bibinfo{year}{2005}\natexlab{}.
\newblock \showarticletitle{Automatic Diagnosis of Performance Problems in Database Management Systems}. In \bibinfo{booktitle}{\emph{{ICAC}}}. \bibinfo{publisher}{{IEEE} Computer Society}, \bibinfo{pages}{326--327}.
\newblock


\bibitem[\protect\citeauthoryear{Chen, Fan, Wu, Tang, Deng, Wang, Li, Tan, Li, Zhou, and Du}{Chen et~al\mbox{.}}{2025}]%
        {Andromeda}
\bibfield{author}{\bibinfo{person}{Sibei Chen}, \bibinfo{person}{Ju Fan}, \bibinfo{person}{Bin Wu}, \bibinfo{person}{Nan Tang}, \bibinfo{person}{Chao Deng}, \bibinfo{person}{Pengyi Wang}, \bibinfo{person}{Ye Li}, \bibinfo{person}{Jian Tan}, \bibinfo{person}{Feifei Li}, \bibinfo{person}{Jingren Zhou}, {and} \bibinfo{person}{Xiaoyong Du}.} \bibinfo{year}{2025}\natexlab{}.
\newblock \showarticletitle{Automatic Database Configuration Debugging using Retrieval-Augmented Language Models}.
\newblock \bibinfo{journal}{\emph{Proc. {ACM} Manag. Data}} \bibinfo{volume}{3}, \bibinfo{number}{1} (\bibinfo{year}{2025}), \bibinfo{pages}{13:1--13:27}.
\newblock


\bibitem[\protect\citeauthoryear{Das, Grbic, Ilic, Jovandic, Jovanovic, Narasayya, Radulovic, Stikic, Xu, and Chaudhuri}{Das et~al\mbox{.}}{2019}]%
        {Azure}
\bibfield{author}{\bibinfo{person}{Sudipto Das}, \bibinfo{person}{Miroslav Grbic}, \bibinfo{person}{Igor Ilic}, \bibinfo{person}{Isidora Jovandic}, \bibinfo{person}{Andrija Jovanovic}, \bibinfo{person}{Vivek~R. Narasayya}, \bibinfo{person}{Miodrag Radulovic}, \bibinfo{person}{Maja Stikic}, \bibinfo{person}{Gaoxiang Xu}, {and} \bibinfo{person}{Surajit Chaudhuri}.} \bibinfo{year}{2019}\natexlab{}.
\newblock \showarticletitle{Automatically Indexing Millions of Databases in Microsoft Azure {SQL} Database}. In \bibinfo{booktitle}{\emph{{SIGMOD} Conference}}. \bibinfo{pages}{666--679}.
\newblock


\bibitem[\protect\citeauthoryear{DeepSeek-AI}{DeepSeek-AI}{2024}]%
        {DeepSeekV3}
\bibfield{author}{\bibinfo{person}{DeepSeek-AI}.} \bibinfo{year}{2024}\natexlab{}.
\newblock \bibinfo{title}{DeepSeek-V3 Technical Report}.
\newblock
\newblock
\showeprint[arxiv]{2412.19437}~[cs.CL]
\urldef\tempurl%
\url{https://arxiv.org/abs/2412.19437}
\showURL{%
\tempurl}


\bibitem[\protect\citeauthoryear{DeepSeek-AI}{DeepSeek-AI}{2025}]%
        {DeepSeekR1}
\bibfield{author}{\bibinfo{person}{DeepSeek-AI}.} \bibinfo{year}{2025}\natexlab{}.
\newblock \bibinfo{title}{DeepSeek-R1: Incentivizing Reasoning Capability in LLMs via Reinforcement Learning}.
\newblock
\newblock
\showeprint[arxiv]{2501.12948}~[cs.CL]
\urldef\tempurl%
\url{https://arxiv.org/abs/2501.12948}
\showURL{%
\tempurl}


\bibitem[\protect\citeauthoryear{Dias, Ramacher, Shaft, Venkataramani, and Wood}{Dias et~al\mbox{.}}{2005}]%
        {ADDM}
\bibfield{author}{\bibinfo{person}{Karl Dias}, \bibinfo{person}{Mark Ramacher}, \bibinfo{person}{Uri Shaft}, \bibinfo{person}{Venkateshwaran Venkataramani}, {and} \bibinfo{person}{Graham Wood}.} \bibinfo{year}{2005}\natexlab{}.
\newblock \showarticletitle{Automatic Performance Diagnosis and Tuning in Oracle}. In \bibinfo{booktitle}{\emph{{CIDR}}}. \bibinfo{publisher}{www.cidrdb.org}, \bibinfo{pages}{84--94}.
\newblock


\bibitem[\protect\citeauthoryear{Dundjerski and Tomasevic}{Dundjerski and Tomasevic}{2022}]%
        {ATDS}
\bibfield{author}{\bibinfo{person}{Dejan Dundjerski} {and} \bibinfo{person}{Milo Tomasevic}.} \bibinfo{year}{2022}\natexlab{}.
\newblock \showarticletitle{Automatic Database Troubleshooting of Azure {SQL} Databases}.
\newblock \bibinfo{journal}{\emph{{IEEE} Trans. Cloud Comput.}} \bibinfo{volume}{10}, \bibinfo{number}{3} (\bibinfo{year}{2022}), \bibinfo{pages}{1604--1619}.
\newblock


\bibitem[\protect\citeauthoryear{Haldar and Mukhopadhyay}{Haldar and Mukhopadhyay}{2011}]%
        {LevenshteinDistance}
\bibfield{author}{\bibinfo{person}{Rishin Haldar} {and} \bibinfo{person}{Debajyoti Mukhopadhyay}.} \bibinfo{year}{2011}\natexlab{}.
\newblock \showarticletitle{Levenshtein Distance Technique in Dictionary Lookup Methods: An Improved Approach}.
\newblock \bibinfo{journal}{\emph{CoRR}}  \bibinfo{volume}{abs/1101.1232} (\bibinfo{year}{2011}).
\newblock


\bibitem[\protect\citeauthoryear{Hirt}{Hirt}{2007}]%
        {downtime}
\bibfield{author}{\bibinfo{person}{Allan Hirt}.} \bibinfo{year}{2007}\natexlab{}.
\newblock \bibinfo{booktitle}{\emph{Pro SQL server 2005 high availability}}.
\newblock \bibinfo{publisher}{Apress}.
\newblock
\urldef\tempurl%
\url{https://www.sqlservercentral.com/wp-content/uploads/2019/05/Hirt_BusinessofAvailability_Apress_780X.pdf}
\showURL{%
\tempurl}


\bibitem[\protect\citeauthoryear{Ilyas, Rekatsinas, Konda, Pound, Qi, and Soliman}{Ilyas et~al\mbox{.}}{2022}]%
        {LLMKGConstruction1}
\bibfield{author}{\bibinfo{person}{Ihab~F. Ilyas}, \bibinfo{person}{Theodoros Rekatsinas}, \bibinfo{person}{Vishnu Konda}, \bibinfo{person}{Jeffrey Pound}, \bibinfo{person}{Xiaoguang Qi}, {and} \bibinfo{person}{Mohamed~A. Soliman}.} \bibinfo{year}{2022}\natexlab{}.
\newblock \showarticletitle{Saga: {A} Platform for Continuous Construction and Serving of Knowledge at Scale}. In \bibinfo{booktitle}{\emph{{SIGMOD} Conference}}. \bibinfo{publisher}{{ACM}}, \bibinfo{pages}{2259--2272}.
\newblock


\bibitem[\protect\citeauthoryear{Jalilifard, Carid{\'{a}}, Mansano, and Cristo}{Jalilifard et~al\mbox{.}}{2020}]%
        {TFIDF}
\bibfield{author}{\bibinfo{person}{Amir Jalilifard}, \bibinfo{person}{Vinicius~Fernandes Carid{\'{a}}}, \bibinfo{person}{Alex Mansano}, {and} \bibinfo{person}{Rogers Cristo}.} \bibinfo{year}{2020}\natexlab{}.
\newblock \showarticletitle{Semantic Sensitive {TF-IDF} to Determine Word Relevance in Documents}.
\newblock \bibinfo{journal}{\emph{CoRR}}  \bibinfo{volume}{abs/2001.09896} (\bibinfo{year}{2020}).
\newblock


\bibitem[\protect\citeauthoryear{Kalmegh, Babu, and Roy}{Kalmegh et~al\mbox{.}}{2017}]%
        {Time1}
\bibfield{author}{\bibinfo{person}{Prajakta Kalmegh}, \bibinfo{person}{Shivnath Babu}, {and} \bibinfo{person}{Sudeepa Roy}.} \bibinfo{year}{2017}\natexlab{}.
\newblock \showarticletitle{Analyzing Query Performance and Attributing Blame for Contentions in a Cluster Computing Framework}.
\newblock \bibinfo{journal}{\emph{CoRR}}  \bibinfo{volume}{abs/1708.08435} (\bibinfo{year}{2017}).
\newblock


\bibitem[\protect\citeauthoryear{Kalmegh, Babu, and Roy}{Kalmegh et~al\mbox{.}}{2019}]%
        {Time2}
\bibfield{author}{\bibinfo{person}{Prajakta Kalmegh}, \bibinfo{person}{Shivnath Babu}, {and} \bibinfo{person}{Sudeepa Roy}.} \bibinfo{year}{2019}\natexlab{}.
\newblock \showarticletitle{iQCAR: inter-Query Contention Analyzer for Data Analytics Frameworks}. In \bibinfo{booktitle}{\emph{{SIGMOD} Conference}}. \bibinfo{publisher}{{ACM}}, \bibinfo{pages}{918--935}.
\newblock


\bibitem[\protect\citeauthoryear{LeCun, Bengio, and Hinton}{LeCun et~al\mbox{.}}{2015}]%
        {DeepLearning}
\bibfield{author}{\bibinfo{person}{Yann LeCun}, \bibinfo{person}{Yoshua Bengio}, {and} \bibinfo{person}{Geoffrey~E. Hinton}.} \bibinfo{year}{2015}\natexlab{}.
\newblock \showarticletitle{Deep learning}.
\newblock \bibinfo{journal}{\emph{Nat.}} \bibinfo{volume}{521}, \bibinfo{number}{7553} (\bibinfo{year}{2015}), \bibinfo{pages}{436--444}.
\newblock


\bibitem[\protect\citeauthoryear{Liu, Yin, Zhao, Ge, Chen, Gao, Li, Wang, Liang, Tan, and Li}{Liu et~al\mbox{.}}{2022}]%
        {PinSQL}
\bibfield{author}{\bibinfo{person}{Xiaoze Liu}, \bibinfo{person}{Zheng Yin}, \bibinfo{person}{Chao Zhao}, \bibinfo{person}{Congcong Ge}, \bibinfo{person}{Lu Chen}, \bibinfo{person}{Yunjun Gao}, \bibinfo{person}{Dimeng Li}, \bibinfo{person}{Ziting Wang}, \bibinfo{person}{Gaozhong Liang}, \bibinfo{person}{Jian Tan}, {and} \bibinfo{person}{Feifei Li}.} \bibinfo{year}{2022}\natexlab{}.
\newblock \showarticletitle{PinSQL: Pinpoint Root Cause SQLs to Resolve Performance Issues in Cloud Databases}. In \bibinfo{booktitle}{\emph{{ICDE}}}. \bibinfo{publisher}{{IEEE}}, \bibinfo{pages}{2549--2561}.
\newblock


\bibitem[\protect\citeauthoryear{Lu, Xie, Li, Li, Nie, Zhao, Yu, Zhang, Sui, Zhu, and Pei}{Lu et~al\mbox{.}}{2022}]%
        {CauseRank}
\bibfield{author}{\bibinfo{person}{Xianglin Lu}, \bibinfo{person}{Zhe Xie}, \bibinfo{person}{Zeyan Li}, \bibinfo{person}{Mingjie Li}, \bibinfo{person}{Xiaohui Nie}, \bibinfo{person}{Nengwen Zhao}, \bibinfo{person}{Qingyang Yu}, \bibinfo{person}{Shenglin Zhang}, \bibinfo{person}{Kaixin Sui}, \bibinfo{person}{Lin Zhu}, {and} \bibinfo{person}{Dan Pei}.} \bibinfo{year}{2022}\natexlab{}.
\newblock \showarticletitle{Generic and Robust Performance Diagnosis via Causal Inference for {OLTP} Database Systems}. In \bibinfo{booktitle}{\emph{{CCGRID}}}. \bibinfo{publisher}{{IEEE}}, \bibinfo{pages}{655--664}.
\newblock


\bibitem[\protect\citeauthoryear{Ma, Yin, Zhang, Wang, Zheng, Jiang, Hu, Luo, Li, Qiu, Li, Chen, and Pei}{Ma et~al\mbox{.}}{2020}]%
        {iSQUAD}
\bibfield{author}{\bibinfo{person}{Minghua Ma}, \bibinfo{person}{Zheng Yin}, \bibinfo{person}{Shenglin Zhang}, \bibinfo{person}{Sheng Wang}, \bibinfo{person}{Christopher Zheng}, \bibinfo{person}{Xinhao Jiang}, \bibinfo{person}{Hanwen Hu}, \bibinfo{person}{Cheng Luo}, \bibinfo{person}{Yilin Li}, \bibinfo{person}{Nengjun Qiu}, \bibinfo{person}{Feifei Li}, \bibinfo{person}{Changcheng Chen}, {and} \bibinfo{person}{Dan Pei}.} \bibinfo{year}{2020}\natexlab{}.
\newblock \showarticletitle{Diagnosing Root Causes of Intermittent Slow Queries in Large-Scale Cloud Databases}.
\newblock \bibinfo{journal}{\emph{Proc. {VLDB} Endow.}} \bibinfo{volume}{13}, \bibinfo{number}{8} (\bibinfo{year}{2020}), \bibinfo{pages}{1176--1189}.
\newblock


\bibitem[\protect\citeauthoryear{Mou, Liu, Sowe, Collarana, and Decker}{Mou et~al\mbox{.}}{2024}]%
        {LLMKGConstruction2}
\bibfield{author}{\bibinfo{person}{Yongli Mou}, \bibinfo{person}{Li Liu}, \bibinfo{person}{Sulayman~K. Sowe}, \bibinfo{person}{Diego Collarana}, {and} \bibinfo{person}{Stefan Decker}.} \bibinfo{year}{2024}\natexlab{}.
\newblock \showarticletitle{Leveraging LLMs Few-shot Learning to Improve Instruction-driven Knowledge Graph Construction}. In \bibinfo{booktitle}{\emph{{VLDB} Workshops}}. \bibinfo{publisher}{VLDB.org}.
\newblock


\bibitem[\protect\citeauthoryear{Ouyang, Zhang, Cheng, Shu, Guo, Yang, Wen, Fan, and Jensen}{Ouyang et~al\mbox{.}}{2025}]%
        {RCRank}
\bibfield{author}{\bibinfo{person}{Biao Ouyang}, \bibinfo{person}{Yingying Zhang}, \bibinfo{person}{Hanyin Cheng}, \bibinfo{person}{Yang Shu}, \bibinfo{person}{Chenjuan Guo}, \bibinfo{person}{Bin Yang}, \bibinfo{person}{Qingsong Wen}, \bibinfo{person}{Lunting Fan}, {and} \bibinfo{person}{Christian~S. Jensen}.} \bibinfo{year}{2025}\natexlab{}.
\newblock \showarticletitle{RCRank: Multimodal Ranking of Root Causes of Slow Queries in Cloud Database Systems}.
\newblock \bibinfo{journal}{\emph{CoRR}}  \bibinfo{volume}{abs/2503.04252} (\bibinfo{year}{2025}).
\newblock


\bibitem[\protect\citeauthoryear{Singh, Vaidya, Kumar, Khosla, Narayanaswamy, Gangadharaiah, and Kraska}{Singh et~al\mbox{.}}{2024}]%
        {panda}
\bibfield{author}{\bibinfo{person}{Vikramank~Y. Singh}, \bibinfo{person}{Kapil Vaidya}, \bibinfo{person}{Vinayshekhar~Bannihatti Kumar}, \bibinfo{person}{Sopan Khosla}, \bibinfo{person}{Balakrishnan Narayanaswamy}, \bibinfo{person}{Rashmi Gangadharaiah}, {and} \bibinfo{person}{Tim Kraska}.} \bibinfo{year}{2024}\natexlab{}.
\newblock \showarticletitle{Panda: Performance Debugging for Databases using {LLM} Agents}. In \bibinfo{booktitle}{\emph{{CIDR}}}. \bibinfo{publisher}{www.cidrdb.org}.
\newblock


\bibitem[\protect\citeauthoryear{Yoon, Niu, and Mozafari}{Yoon et~al\mbox{.}}{2016}]%
        {DBSherlock}
\bibfield{author}{\bibinfo{person}{Dong~Young Yoon}, \bibinfo{person}{Ning Niu}, {and} \bibinfo{person}{Barzan Mozafari}.} \bibinfo{year}{2016}\natexlab{}.
\newblock \showarticletitle{DBSherlock: {A} Performance Diagnostic Tool for Transactional Databases}. In \bibinfo{booktitle}{\emph{{SIGMOD} Conference}}. \bibinfo{publisher}{{ACM}}, \bibinfo{pages}{1599--1614}.
\newblock


\bibitem[\protect\citeauthoryear{Zeng and Klabjan}{Zeng and Klabjan}{2019}]%
        {volatility}
\bibfield{author}{\bibinfo{person}{Yaxiong Zeng} {and} \bibinfo{person}{Diego Klabjan}.} \bibinfo{year}{2019}\natexlab{}.
\newblock \showarticletitle{Online adaptive machine learning based algorithm for implied volatility surface modeling}.
\newblock \bibinfo{journal}{\emph{Knowl. Based Syst.}}  \bibinfo{volume}{163} (\bibinfo{year}{2019}), \bibinfo{pages}{376--391}.
\newblock


\bibitem[\protect\citeauthoryear{Zhong, Wu, Li, Peng, and Wu}{Zhong et~al\mbox{.}}{2024}]%
        {KGSurvey}
\bibfield{author}{\bibinfo{person}{Lingfeng Zhong}, \bibinfo{person}{Jia Wu}, \bibinfo{person}{Qian Li}, \bibinfo{person}{Hao Peng}, {and} \bibinfo{person}{Xindong Wu}.} \bibinfo{year}{2024}\natexlab{}.
\newblock \showarticletitle{A Comprehensive Survey on Automatic Knowledge Graph Construction}.
\newblock \bibinfo{journal}{\emph{{ACM} Comput. Surv.}} \bibinfo{volume}{56}, \bibinfo{number}{4} (\bibinfo{year}{2024}), \bibinfo{pages}{94:1--94:62}.
\newblock


\bibitem[\protect\citeauthoryear{Zhou, Gao, Zhou, and Li}{Zhou et~al\mbox{.}}{2025a}]%
        {zhou2025cracksql}
\bibfield{author}{\bibinfo{person}{Wei Zhou}, \bibinfo{person}{Yuyang Gao}, \bibinfo{person}{Xuanhe Zhou}, {and} \bibinfo{person}{Guoliang Li}.} \bibinfo{year}{2025}\natexlab{a}.
\newblock \showarticletitle{{Cracking SQL Barriers:} {An} LLM-based Dialect Transaltion System}.
\newblock \bibinfo{journal}{\emph{Proc. {ACM} Manag. Data}} \bibinfo{volume}{3}, \bibinfo{number}{3 (SIGMOD)} (\bibinfo{year}{2025}).
\newblock


\bibitem[\protect\citeauthoryear{Zhou, Lin, Zhou, and Li}{Zhou et~al\mbox{.}}{2024b}]%
        {BID}
\bibfield{author}{\bibinfo{person}{Wei Zhou}, \bibinfo{person}{Chen Lin}, \bibinfo{person}{Xuanhe Zhou}, {and} \bibinfo{person}{Guoliang Li}.} \bibinfo{year}{2024}\natexlab{b}.
\newblock \showarticletitle{Breaking It Down: An In-depth Study of Index Advisors}.
\newblock \bibinfo{journal}{\emph{Proc. {VLDB} Endow.}} \bibinfo{volume}{17}, \bibinfo{number}{10} (\bibinfo{year}{2024}), \bibinfo{pages}{2405--2418}.
\newblock


\bibitem[\protect\citeauthoryear{Zhou, Lin, Zhou, Li, and Wang}{Zhou et~al\mbox{.}}{2024c}]%
        {TRAP}
\bibfield{author}{\bibinfo{person}{Wei Zhou}, \bibinfo{person}{Chen Lin}, \bibinfo{person}{Xuanhe Zhou}, \bibinfo{person}{Guoliang Li}, {and} \bibinfo{person}{Tianqing Wang}.} \bibinfo{year}{2024}\natexlab{c}.
\newblock \showarticletitle{TRAP: Tailored Robustness Assessement for Index Advisors via Adversarial Perturbation}. In \bibinfo{booktitle}{\emph{{ICDE}}}. \bibinfo{pages}{to appear}.
\newblock


\bibitem[\protect\citeauthoryear{Zhou, Sun, Zhou, Li, Liu, Wu, and Wang}{Zhou et~al\mbox{.}}{2025c}]%
        {gaussmaster}
\bibfield{author}{\bibinfo{person}{Wei Zhou}, \bibinfo{person}{Ji Sun}, \bibinfo{person}{Xuanhe Zhou}, \bibinfo{person}{Guoliang Li}, \bibinfo{person}{Luyang Liu}, \bibinfo{person}{Hao Wu}, {and} \bibinfo{person}{Tianyuan Wang}.} \bibinfo{year}{2025}\natexlab{c}.
\newblock \showarticletitle{GaussMaster: An LLM-based Database Copilot System}.
\newblock \bibinfo{journal}{\emph{CoRR}}  \bibinfo{volume}{abs/2506.23322} (\bibinfo{year}{2025}).
\newblock


\bibitem[\protect\citeauthoryear{Zhou, He, Zhou, Chen, Tang, Zhao, Tong, Li, Chen, Zhou, Sun, Hui, Wang, He, Liu, Zhou, and Wu}{Zhou et~al\mbox{.}}{2025b}]%
        {OurLLMSurvey}
\bibfield{author}{\bibinfo{person}{Xuanhe Zhou}, \bibinfo{person}{Junxuan He}, \bibinfo{person}{Wei Zhou}, \bibinfo{person}{Haodong Chen}, \bibinfo{person}{Zirui Tang}, \bibinfo{person}{Haoyu Zhao}, \bibinfo{person}{Xin Tong}, \bibinfo{person}{Guoliang Li}, \bibinfo{person}{Youmin Chen}, \bibinfo{person}{Jun Zhou}, \bibinfo{person}{Zhaojun Sun}, \bibinfo{person}{Binyuan Hui}, \bibinfo{person}{Shuo Wang}, \bibinfo{person}{Conghui He}, \bibinfo{person}{Zhiyuan Liu}, \bibinfo{person}{Jingren Zhou}, {and} \bibinfo{person}{Fan Wu}.} \bibinfo{year}{2025}\natexlab{b}.
\newblock \showarticletitle{A Survey of {LLM} × {DATA}}.
\newblock \bibinfo{journal}{\emph{arXiv Preprint}} (\bibinfo{year}{2025}).
\newblock
\urldef\tempurl%
\url{https://arxiv.org/abs/2505.18458}
\showURL{%
\tempurl}


\bibitem[\protect\citeauthoryear{Zhou, Jin, Sun, Zhao, Yu, Li, Wang, Li, and Liu}{Zhou et~al\mbox{.}}{2021}]%
        {DBMind}
\bibfield{author}{\bibinfo{person}{Xuanhe Zhou}, \bibinfo{person}{Lianyuan Jin}, \bibinfo{person}{Ji Sun}, \bibinfo{person}{Xinyang Zhao}, \bibinfo{person}{Xiang Yu}, \bibinfo{person}{Shifu Li}, \bibinfo{person}{Tianqing Wang}, \bibinfo{person}{Kun Li}, {and} \bibinfo{person}{Luyang Liu}.} \bibinfo{year}{2021}\natexlab{}.
\newblock \showarticletitle{DBMind: {A} Self-Driving Platform in openGauss}.
\newblock \bibinfo{journal}{\emph{Proc. {VLDB} Endow.}} \bibinfo{volume}{14}, \bibinfo{number}{12} (\bibinfo{year}{2021}), \bibinfo{pages}{2743--2746}.
\newblock


\bibitem[\protect\citeauthoryear{Zhou, Li, Sun, Liu, Chen, Wu, Liu, Feng, and Zeng}{Zhou et~al\mbox{.}}{2024a}]%
        {dbot2024}
\bibfield{author}{\bibinfo{person}{Xuanhe Zhou}, \bibinfo{person}{Guoliang Li}, \bibinfo{person}{Zhaoyan Sun}, \bibinfo{person}{Zhiyuan Liu}, \bibinfo{person}{Weize Chen}, \bibinfo{person}{Jianming Wu}, \bibinfo{person}{Jiesi Liu}, \bibinfo{person}{Ruohang Feng}, {and} \bibinfo{person}{Guoyang Zeng}.} \bibinfo{year}{2024}\natexlab{a}.
\newblock \showarticletitle{D-Bot: Database Diagnosis System using Large Language Models}.
\newblock \bibinfo{journal}{\emph{Proc. {VLDB} Endow.}} \bibinfo{volume}{17}, \bibinfo{number}{10} (\bibinfo{year}{2024}), \bibinfo{pages}{2514--2527}.
\newblock


\end{thebibliography}
